\documentclass[fleqn,usenatbib]{mnras}

\usepackage{newtxtext,newtxmath}

\usepackage[T1]{fontenc}

\DeclareRobustCommand{\VAN}[3]{#2}
\let\VANthebibliography\thebibliography
\def\thebibliography{\DeclareRobustCommand{\VAN}[3]{##3}\VANthebibliography}

\usepackage{graphicx}	
\usepackage{amsmath}	
\usepackage{booktabs}
\usepackage{tablefootnote}

\newcommand{\orcid}[1]{\href{https://orcid.org/#1}{\includegraphics[width=10pt]{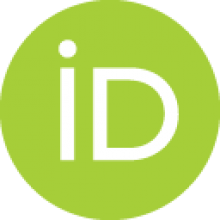}}}

\newcommand{\citnp}[1]{\textcolor{blue}{#1} (\textcolor{blue}{in prep})}

\defcitealias{Jarrett2023}{J23}
\defcitealias{Cluver2014}{C14}
\defcitealias{Ramakrishnan2023}{R23}

\title[WISE2MBH Algorithm]{WISE2MBH: A scaling-based algorithm for probing supermassive black hole masses through WISE catalogues}

\author[J. Hernández-Yévenes et al.]{
J. Hernández-Yévenes\ \orcid{0000-0001-5845-7538},$^{1}$\thanks{Contact e-mail: \href{mailto:jheryev@gmail.com}{jheryev@gmail.com}}
N. Nagar\ \orcid{0000-0001-6920-662X},$^{1}$
V. Arratia\ \orcid{0000-0003-4785-2297}$^{1}$
and T. H. Jarrett\ \orcid{0000-0002-4939-734X}$^{2,3,4}$
\\
$^1$Departamento de Astronomía, Universidad de Concepción, Av. Esteban Iturra s/n, Casilla 160-C, Concepción, Chile\\
$^2$Department of Astronomy, University of Cape Town, Rondebosch, 7700, South Africa\\
$^3$Centre for Astrophysics and Supercomputing, Swinburne University of Technology, John Street, Hawthorn, 3122, Australia\\
$^4$Institute for Astronomy, University of Hawaii at Hilo, 640 N Aohoku Pl 209, Hilo, HI 96720, USA}

\date{Accepted 2024 May 15. Received 2024 May 14; in original form 2023 September 27}

\pubyear{2024}

\begin{document}
\label{firstpage}
\pagerange{\pageref{firstpage}--\pageref{lastpage}}
\maketitle

\begin{abstract}
Supermassive Black Holes (SMBHs) are commonly found at the centers of massive galaxies. Estimating their masses ($M_\text{BH}$) is crucial for understanding galaxy-SMBH co-evolution. We present WISE2MBH, an efficient algorithm that uses cataloged Wide-field Infrared Survey Explorer (WISE) magnitudes to estimate total stellar mass ($M_*$) and scale this to bulge mass ($M_\text{Bulge}$), and $M_\text{BH}$, estimating the morphological type ($T_\text{Type}$) and bulge fraction ($B/T$) in the process. WISE2MBH uses scaling relations from the literature or developed in this work, providing a streamlined approach to derive these parameters. It also distinguishes QSOs from galaxies and estimates the galaxy $T_\text{Type}$ using WISE colors with a relation trained with galaxies from the 2MASS Redshift Survey.
WISE2MBH performs well up to $z\sim0.5$ thanks to K-corrections in magnitudes and colors. WISE2MBH $M_\text{BH}$ estimates agree very well with those of a selected sample of local galaxies with $M_\text{BH}$ measurements or reliable estimates: a Spearman score of $\sim$0.8 and a RMSE of $\sim$0.63 were obtained. When applied to the ETHER sample at $z\leq0.5$, WISE2MBH provides $\sim$1.9 million $M_\text{BH}$ estimates (78.5\% new) and $\sim$100 thousand upper limits. The derived local black hole mass function (BHMF) is in good agreement with existing literature BHMFs. Galaxy demographic projects, including target selection for the Event Horizon Telescope, can benefit from WISE2MBH for up-to-date galaxy parameters and $M_\text{BH}$ estimates. The WISE2MBH algorithm is publicly available on GitHub.
\end{abstract}



\section{Introduction}
Supermassive black holes (SMBH) are characterized by having masses ($M_\text{BH}$) ranging from $\sim$ $10^5\text{ to }10^{10} M_\odot$ and are believed to be located at the centers of all galaxies with a bulge, including the Milky Way \citep[e.g.,][]{Ferrarese2005, Graham2016}. The presence of SMBH is inferred from observations of stellar and gas motions in galactic nuclei \citep[e.g.,][]{Genzel2010,Saglia2016}, as well as powerful nuclear X-ray to radio emission \citep[e.g.,][]{Broderick2015, Liu2022}.
The Event Horizon Telescope (EHT) has provided the most direct evidence of the existence of SMBH \citep{EHT_m87_1, EHT_sgra_1}. SMBH play a crucial role in shaping the evolution and structure of galaxies, as they can affect the surrounding stars and gas through `feedback', and they are expected to co-evolve with their host galaxies \citep{Kormendy2013}.

\begin{figure*}
    \centering
    \includegraphics[width=\textwidth]{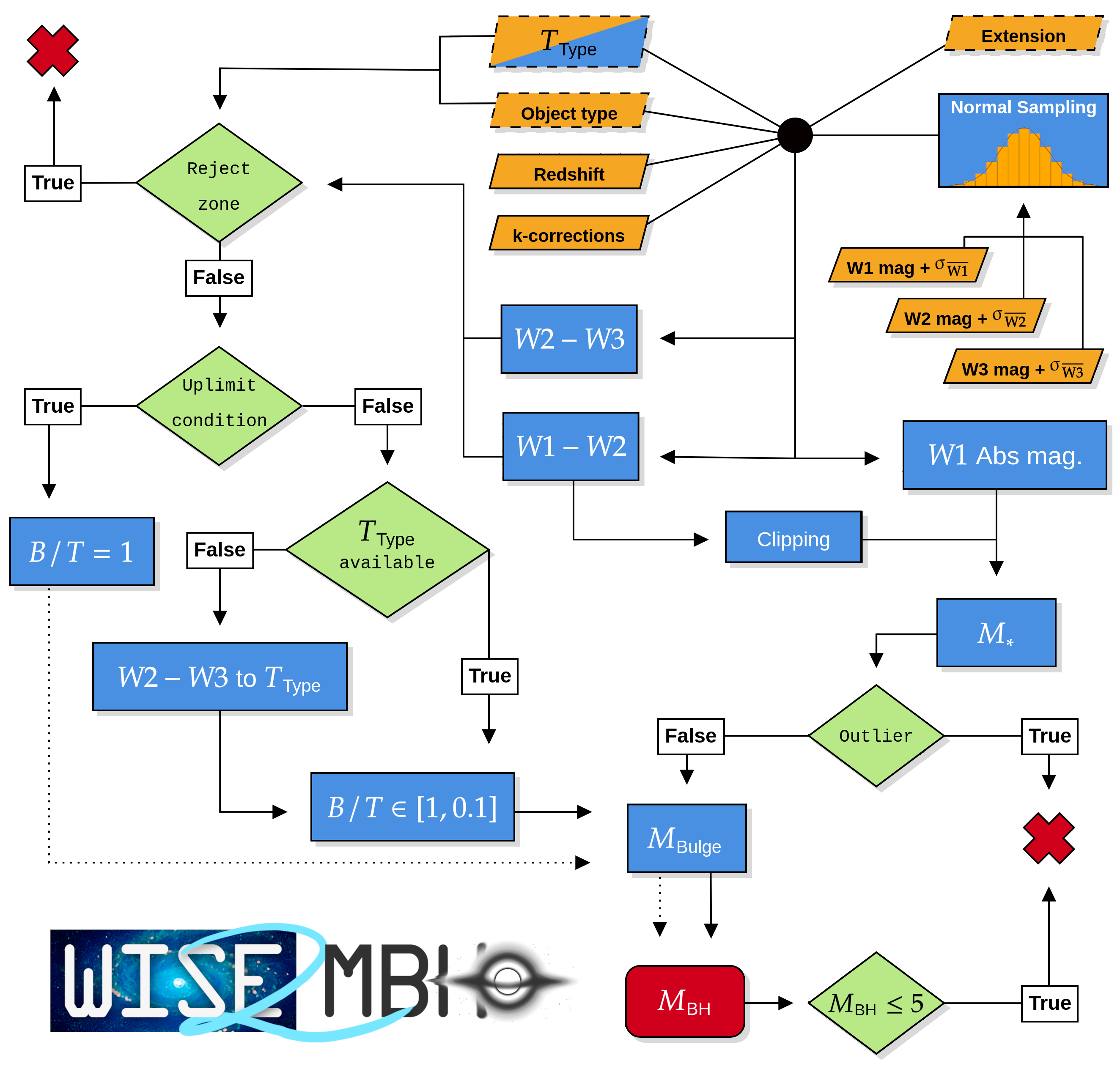}
    \caption{WISE2MBH: A simple algorithm that makes use of WISE cataloged data and a spectroscopic redshift to estimate the stellar mass, morphological type, bulge fraction, and $M_\text{BH}$ of an extragalactic source. Solid (dotted) lines represent the main path to estimate a value (or upper limit), or to reject an object from the algorithm.
    Orange (blue) boxes show the input (derived) quantities; boxes with both colors can either be provided to or are estimated by the algorithm. Inputs in dashed boxes are optional. WISE magnitudes with their respective mean photometric errors ($\upsigma_{\overline{WX}}$) are used to generate random normal samples of size $10^4$ for a Monte Carlo approach to error propagation.}
    \label{fig:flowchart}
\end{figure*}

Active galactic nuclei (AGN) are a manifestation of SMBH that are powered by a luminous accretion process at the centers of numerous galaxies. AGN emission can cover the entire electromagnetic spectrum \citep{Padovani2017}. 
At infrared (IR) wavelengths, the emission is primarily attributed to a toroidal arrangement of dust, which absorbs the radiation emitted by the central accretion disk and re-radiates it at IR \citep{Netzer2015, Hickox2018}.

Quasars (QSO) are a type of extremely luminous AGN (with bolometric luminosity, $\log L_\text{bol}$,  in the range of $44\text{ to }48\text{ erg s}^{-1}$) that are powered by high accretion rates onto SMBH \citep[$L_\text{bol}/L_\text{Edd}\sim10^{-2.9}$\text{ to }$10^{1.8}$,][]{Kong2018}. Given these high accretion rates, QSOs are among the most luminous objects in the universe, making it difficult (though not completely impossible) to discriminate the morphology of their host galaxies at any redshift \citep{Dunlop2003}. QSOs are among the earliest and most distant observable objects \citep[e.g., $z \sim$7.64,][]{Wang2021}, and are important probes of the early universe and the formation and growth of galaxies and their SMBHs \citep{Inayoshi2020}.

The Wide-field Infrared Survey Explorer \citep[WISE,][]{Wright2010} was a NASA IR$-$Wavelength astronomical space telescope that surveyed the entire sky in four IR bands, with central wavelengths at 3.4 $\mu$m, 4.6 $\mu$m, 12 $\mu$m, and 22 $\mu$m (W1, W2, W3 and W4, respectively). These bands can be used to study several features of a galaxy. The shorter wavelength bands (W1 and W2) predominantly capture emission from stars \citep[e.g.,][]{Jarrett2011,Jarrett2012,Norris2014} and warm dust \citep[e.g.,][]{Lyu2019, Noda2020, Li2023}, providing insights into stellar populations in galaxies \citep[e.g.,][]{Kettlety2018}. The longer wavelength bands (W3 and W4) are sensitive to emission from cooler dust, revealing regions of colder dust associated with older stars \citep[e.g.,][]{Singh2021,Li2023_2}, allow estimation of star formation rates \citep[SFRs, e.g.,][]{Lee2013, Cluver2017} and potentially highlighting the presence of AGN \citep[e.g.,][]{Lyu2022, Hviding2022}. By analyzing the relative intensities and the spatial distribution of emission across these bands, it is possible to obtain the temperature, composition, and distribution of a galaxy's dust and stars, thereby unraveling its evolutionary history and physical characteristics.

One of the most relevant physical characteristics that can be estimated with WISE photometry is the total stellar mass of a galaxy ($M_*$). As stated by \citet[hereafter \citetalias{Cluver2014}]{Cluver2014} the W1 band is the most sensitive to the light emitted by the bulk of stellar population in galaxies \citep[e.g.,][]{Meidt2012,Norris2014}, thus allowing a determination of the mass attributed to the mass-dominant stellar population by using a mass-to-light (M/L) ratio \citep[e.g.,][]{Kettlety2018}. The M/L ratio can be constrained using the W1$-$W2 color. More recently, \citet[hereafter \citetalias{Jarrett2023}]{Jarrett2023} have refined this process, developing a more stringent method to obtain stellar mass estimates from only the W1 band and assuming a global M/L ratio of $\sim$0.35 for all galaxies. Since early-type galaxies have  higher M/L ($\sim$0.8) ratios, WISE-based estimates obtained using individually tailored M/L ratios give results that are significantly different from the W1-only estimate. In instances of low uncertainty in WISE colors, the M/L ratio can be obtained through the use of the W1$-$W2 or W1$-$W3 color, and hence better correct for the range of M/L that is observed from early to late-types.

\begin{figure*}
\centering
\includegraphics[width=0.48\textwidth]{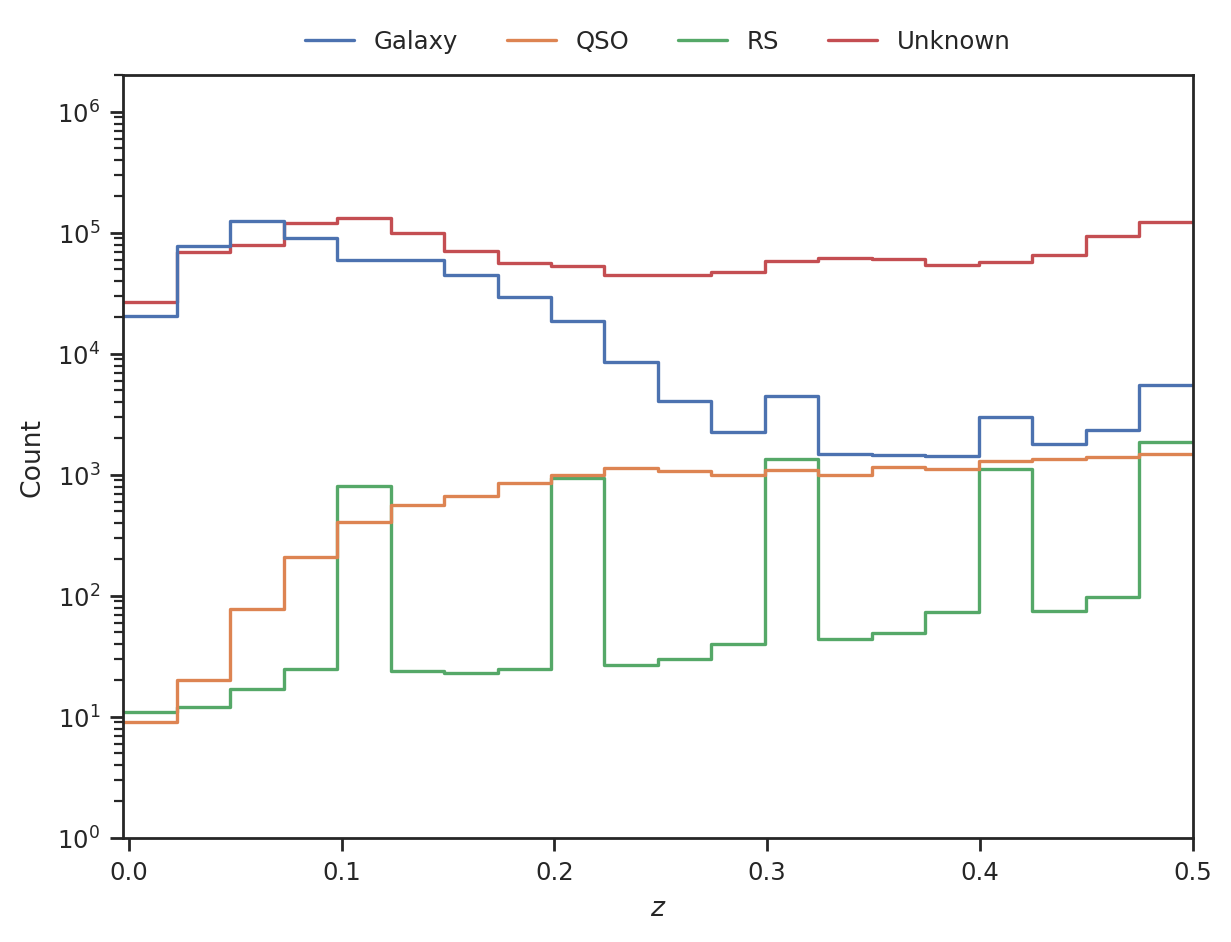}
\includegraphics[width=0.48\textwidth]{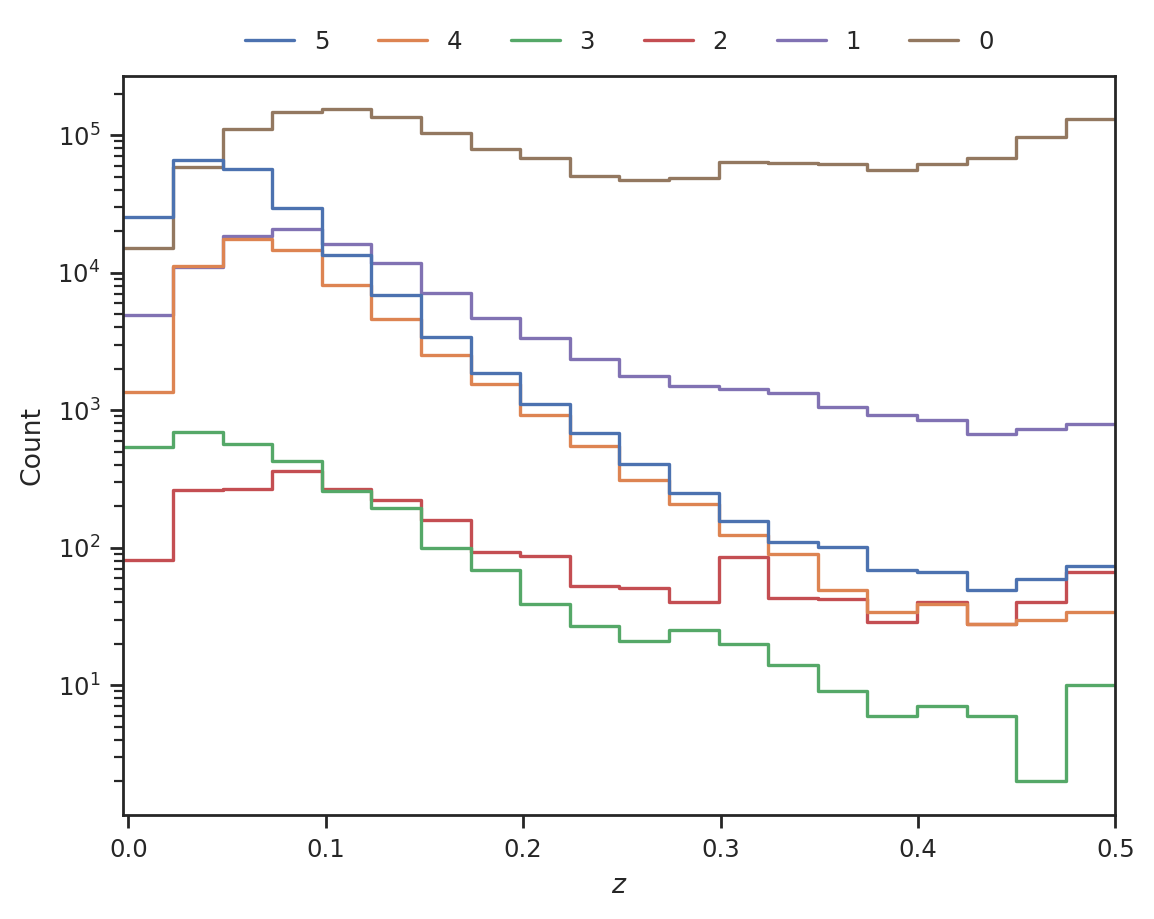}
\caption{Redshift distribution of the WISE2MBH parent sample. Left: Separated by object type. Right: Separated by source extension (as listed in the AllWISE catalog). Galaxies and unknown type sources dominate by a few orders of magnitude over other object types, while point-like (3-0) sources dominate at z $\gtrsim$ 0.1.}
\label{fig:hist_data_input}
\end{figure*}

Morphology and its evolution is another relevant property of galaxies and \citep[e.g.,][]{Abraham1994, Abraham2001, Willett2013}. Many studies \citep[e.g.,][]{Abraham1996,Whyte2002,Pahre2004} suggest that IR morphological classifications of galaxies can be superior to optical classifications, due to the physical properties that can be studied in the mid-IR, e.g., SFR and stellar populations, which evolve with the morphological type of the galaxy. Using WISE, distinct populations of early-type and late-type galaxies shown clearly different IR colors \citep[e.g.,][]{Wright2010,Lee2017,Yao2020}. Recently, \cite{Jarrett2019} showed a Hubble sequence-like morphological evolution with the W2$-$W3 color for a sample of well-known nearby galaxies, where early-type galaxies showed redder (stellar-dominated) colors compared to late-type (ISM + stellar colors) galaxies, with a clear sequence from the early-type to late-type. When combined with $M_*$ estimates (e.g., \citetalias{Cluver2014}, \citetalias{Jarrett2023}), the most massive galaxies were shown to be dominated by high fractions ($\geq0.8$) of spheroid-like galaxies. These authors found the same behavior when exploring the bulge-to-total ratios ($B/T$): galaxies with $B/T$ $\geq0.9$ are dominantly at W2$-$W3 $\leq1.5$, which was set as the cutoff between spheroid galaxies and intermediate disks. Other studies have also shown that early-type and late-type galaxies showed distinct distributions of W2$-$W3 color \citep[e.g.,][]{Sadler2014,Cluver2020}, with the former being once again the most massive galaxies. In the sample of \cite{Cluver2020} at $\log M_*\geq11.2$, the percentage of early-type galaxies reaches $\sim80\%$. 

Powerful AGN are characterized by their high IR luminosity, resulting in strong WISE detections. 
WISE can differentiate between these AGN and galaxies using WISE color-color criteria \citep[e.g.,][]{Jarrett2011, Hviding2022}. This differentiation is important as AGN play a crucial role in the evolution of galaxies: they have the capability to heat the surrounding gas and dust, suppress the formation of new stars, and ultimately quench star formation. 
\cite{Stern2012} proposed a simple W1$-$W2 cutoff to successfully identify AGN to a depth of W2 $\sim 15$. This cutoff was exploited, improved, and used for the creation of the WISE AGN catalog \citep{Assef2018} and to derive new criteria to identify low luminosity AGN \citep[$\log L_\text{bol}$ in the range of $40\text{ to }44\text{ erg s}^{-1}$;][]{Hviding2022}. 

WISE colors are useful in identifying AGN, being notably good for type 2 obscured AGN \citep[e.g.,][]{Hviding2022}, but are poor in identifying relatively weak AGN and highly obscured AGN at modest to large redshifts \citep[z$>2$ e.g.,][]{Hickox2018,Lyu2022R,Lyu2023}. The latter may constitute a significant fraction ($\geq$40\%) of the QSO population at high redshifts \citep{Lyu2023}. For complete identification of AGN, observations at multiple wavelengths are required \citep[e.g.,][]{Yao2022}.
In this work AGN identification is not a goal in itself; it is used to evaluate and correct AGN contamination to WISE W1 and W2 based host galaxy stellar mass estimates. Since highly obscured (and relatively weak) AGN are expected to contribute only a small fraction of the total (AGN plus galaxy) flux at wavelengths shorter than 5\micron\ \citep[e.g.,][Fig. 6]{Lyu2022R} they are not expected to significantly contaminate the host galaxy stellar mass estimates calculated here.

Empirically, a strong correlation is observed between the $M_\text{BH}$ and the bulge mass ($M_\text{Bulge}$) in elliptical galaxies, as well as in spiral galaxies with pseudo-bulges and classical bulges \citep[e.g.,][]{Haring2004,Kormendy2013,Schutte2019} with previous theoretical approaches supporting this idea \citep[e.g.,][]{Croton2006}. Combining this with the observed evolution of $B/T$ with the $T_\text{Type}$ of a galaxy \citep[e.g.,][]{Wang2019,Dimauro2022,Quilley2022}, it is possible to exploit the WISE magnitudes and colors to estimate $M_\text{Bulge}$ and then $M_\text{BH}$, from $M_*$.

Combining all of the above, WISE photometry can be used to distinguish between different types of extragalactic objects, e.g. QSOs, ULIRGS, AGN, and galaxies, and then for galaxies, to estimate the  total stellar mass and morphology \citep[using colors between the W1, W2, and W3 bands, e.g.,][\citetalias{Cluver2014}, \citetalias{Jarrett2023}]{Wright2010} and thus estimate $M_\text{Bulge}$ and $M_\text{BH}$.

In this work, we introduce a new algorithm, WISE2MBH (see Fig. \ref{fig:flowchart}), which takes advantage of existing relationships derived from WISE data, the proportionality between the masses of the galaxy bulge and its SMBH, and new scaling relationships derived here. The WISE2MBH algorithm is capable of classifying regular galaxies, estimating their morphological type, and thus their bulge to total mass ratio, and estimating the mass of the SMBH. Additionally, it can identify QSOs from WISE colors; due to the AGN contamination, the algorithm provides upper limit values for these. This algorithm and the resulting sample of SMBH masses are relevant to the study of individual sources using powerful instruments such as the EHT and its next-generation upgrade \citep[ngEHT,][]{Johnson2023, Doeleman2023}, as well as to studies of SMBH populations and evolution. In Sect. \ref{sec:data} we introduce the data used as input to the algorithm, in Sect. \ref{sec:kcorrections} to \ref{sec:algorithm} we explain, in detail, the main steps of the algorithm. In Sect. \ref{sec:results} we present the main results and statistics of the WISE2MBH final sample generated by the algorithm, in Sect. \ref{sec:discussion} we briefly discuss the results, their relevance, main assumptions and limitations, and lastly in Sect. \ref{sec:conclusion} we present our conclusions.

Throughout this work, we use Vega magnitudes and adopt the cosmological parameters of \cite{Planck2020_6}: $\Omega_\text{m} = 0.31$, $\Omega_\Lambda =0.69$, and $H_0 = 67.66 \text{ km s}^{-1} \text{Mpc}^{-1}$.

\section{Data}\label{sec:data}

\begin{table*}
\centering
\caption{Statistics of the WISE2MBH parent sample.}
\resizebox{\textwidth}{!}{%
\begin{tabular*}{\textwidth}{@{}l@{\hspace*{34pt}}c@{\hspace*{34pt}}c@{\hspace*{34pt}}c@{\hspace*{34pt}}c@{\hspace*{34pt}}c@{\hspace*{34pt}}c@{\hspace*{34pt}}c@{}}
\toprule
Object type  & $N$       & $z$      & $M_\text{BH}$     & \texttt{ex}           & W1            & W2            & W3            \\ \midrule
Galaxy       & 601658    & 99.80    & 60.39             & 30.95 / 69.05         & 99.98 / <0.01 & 99.49 / 0.51  & 69.98 / 29.97 \\
QSO          & 16200     & 99.27    & 55.98             & 0.81 / 99.18          & 99.96 / <0.01 & 99.87 / 0.13  & 95.41 / 4.54 \\
RS           & 9360      & 71.83    & 3.31              & 0.33 / 99.66          & 99.83 / 0.13  & 99.48 / 0.52  & 67.13 / 32.83 \\
Unknown      & 1468364   & 98.85    & 59.40             & 5.95 / 94.05          & 99.81 / 0.15  & 99.40 / 0.60  & 48.91 / 51.05 \\
Total        & 2095582   & 99.01    & 59.41             & 13.06 / 86.94         & 99.86 / 0.12  & 99.43 / 0.57  & 55.40 / 44.56 \\ \midrule
Total ETHER  & 2202339   & --.--    & --.--             & --.-- / --.--         & --.-- / --.-- & --.-- / --.-- & --.-- / --.-- \\ \bottomrule
\end{tabular*}
}
\flushleft\footnotesize{\textit{Notes:} This sample is composed of sources from the ETHER sample crossmatched with both AllWISE and WXSC. From left to right, we list the number of sources, and the percentages of sources with a spectroscopic redshift, existing $M_\text{BH}$ estimate, extension flag (\texttt{ex}) equal to 5-4/3-2-1-0 (completely extended/point-like) and photometric quality flags (\texttt{qph}) in the W1, W2, and W3 WISE bands with \texttt{qph} equal to A-B-C/U.}
\label{table:summary}
\end{table*}

\subsection{ETHER sample}\label{sec:ether}
The Event Horizon and Environs sample (ETHER) aims to be the definitive sample and database from which to choose targets for the EHT and ngEHT. 
The database, its algorithms, and references for data sources, was first presented in 
\citet[hereafter \citetalias{Ramakrishnan2023}]{Ramakrishnan2023}. 
ETHER has since been expanded by including the following literature and database samples:
(a) all galaxies in the HyperLeda database \citep{Makarov2014} with recessional velocity (defined nonrelativistically, i.e., $z = v/c$) less than 100.000 km/s;
(b) the Million Quasar catalog \citep[Milliquas,][]{Flech2023};
(c) the Veron-Cetty and Veron AGN catalog \citep[Version 13,][]{Veroncetty2010} with updates from \citep{Flech2013};
(d) the 2M++ redshift survey \citep{Lavaux2011}; 
(e) the ROMA BZCAT blazar catalog \citep[5th Edition,][]{Massaro2016};
(f) the 2MRS sample \citep{Huchra2012}; and
(g) the $\sim$ million galaxies with SDSS and WISE photometry in the catalog of \citet{Chang2015}. 
Several other individual black hole masses and radio fluxes from the literature have also been incorporated. Full details on the updated ETHER sample will be published in \citnp{Nagar et al.}.

Given the above updates and after consolidating multiple entries from the same source, the ETHER sample currently contains 3.8 million extragalactic sources, of which 233 have $M_\text{BH}$ measurements, including  methods such as stellar dynamics \citep[e.g.,][]{Thater2019}, gas kinematics \citep[e.g.,][]{Boizelle2021}, and reverberation mapping \citep[][]{Bentz2015}, and $\sim$860,000 have $M_\text{BH}$ estimates. Of the estimates, $\sim$331,000 are from the M-sigma relationship, $\sim$525,000 are from `single-epoch' spectroscopy and using scaling relationships from reverberation mapping \citep[e.g.,][]{Dalla2020,Rakshit2021}, $\sim$3,000 are from M-L$_{\rm bulge}$ estimations, and $\sim$600 are from other `fundamental-plane' type relationships (see \citetalias{Ramakrishnan2023} for more details).

Astroqueries to the NASA Extragalactic Database (NED) and SIMBAD are used to incorporate and update positions, spectroscopic redshifts (thus luminosity and angular distances for objects at D $\geq$ 50 Mpc), object types, morphological types ($T_\text{Type}$), AGN classifications, and radio to X-ray fluxes. The object type comes directly from NED source classifications, which have shown great precision ($\geq80\%$) for nearby sources \citep[$D\leq11$ Mpc,][]{Kuhn2022}. 
The object types in ETHER, incorporated from NED, are as follows: `Galaxies' cover regular galaxies over the range from elliptical to spiral galaxies, `QSOs' denote galaxies with significant nuclear activity thus luminosity, `Radio Sources' (RS) refer to sources detected in the radio regime, without any distinction between galaxies or QSOs. If a source lacks any NED classification, it is designated as an `Unknown' object type. Through visual inspection of sub-samples using the Sloan Digital Sky Survey (SDSS), and the color-color criteria using WISE colors (see Section \ref{sec:ident-sources}), we established that this NED source classification was sufficiently accurate for our needs. 

Morphological types ($T_\text{Type}$) are available for a significant fraction of the sample \citep[$22.4\%$ via NED and SIMBAD queries and from individual samples e.g.,][]{Huchra2012, Makarov2014}. These are predominantly  E ($-$6 to $-4$) and Sc (4--5), representing $37.3\%$ and $23.8\%$ of all sources with available morphology, respectively. The large fractions in these two $T_\text{Type}$ bins is primarily due to the binary classification of $T_\text{Type}$ in some ingested samples \citep[e.g.,][]{Dobrycheva2013}. When available, this $T_\text{Type}$ is used as an input to the algorithm; when not available, the $T_\text{Type}$ is estimated from the W2$-$W3 color (see Section \ref{sec:est-ttype}). All ETHER sources at $z\leq0.5$ ($\sim$2.3 M) form the parent sample for this work. 

\subsection{WISE catalogues}\label{sec:allwise}
The WISE mission, funded by NASA as a Medium-Class Explorer mission, features a space-based infrared telescope with megapixel cameras cooled by a two-stage solid hydrogen cryostat. This telescope conducted an all-sky survey which simultaneously captured images in four broad spectral bands: W1, W2, W3, and W4, centered on 3.4, 4.6, 12, and 22 $\mu$m with an angular resolution of 6.1\arcsec, 6.4\arcsec, 6.5\arcsec and 12.0\arcsec, and achieving sensitivities of 0.08, 0.11, 1 and 6 mJy, respectively. In this work, we use both the AllWISE catalog \citep{Cutri2021} and the WISE Extended Source Catalog \citep[WXSC,][]{Jarrett2019}. The latter includes mid-infrared photometry and measured global properties of the 100 largest (in angular size) galaxies in WISE.

\subsubsection{AllWISE catalog}
The AllWISE catalog, which combines data from the WISE cryogenic and NEOWISE \citep{Mainzer2011} catalogues, contains almost 750 million sources, including galaxies, stars, brown dwarfs, and asteroids, making it one of the most comprehensive IR catalogues ever created. 
This catalog provides photometric quality flags (\texttt{qph}) for a given detection for all sources, making it the best option to extract data for our purposes; we accepted only sources with quality A, B, C or U in the first three bands, which translates to detections with signal-to-noise ratios (S/N) in the ranges of $\text{S/N}\geq10$, $10>\text{S/N}\geq3$, $3>\text{S/N}\geq2$ and $\text{S/N}<2$, respectively, with the last flag considered an upper limit with $95\%$ confidence.
Extension flags (\texttt{ex}) are also provided in this catalog, allowing easy differentiation between extended and point sources. This flag is directly related to the 2MASS Extended Source Catalog \citep[XSC, ][]{Jarrett2000}, ranging from 5 for completely extended sources to 0 for point sources. 
This flag contributes to the overall quality flag of our estimates and thus allows the selection of high-quality sub-samples (e.g., using only extended sources) from our overall sample. 
In this work, sources with \texttt{ex} equal to 4 or 5 are considered extended, while those with values 0 to 3 are considered point sources. 
Thus, in our definition, a WISE extended source is one whose position
falls within 5\arcsec of the central position of a 2MASS XSC source, and a WISE point source is one that is not associated with a 2MASS XSC source, or is offset by $\geq5\arcsec$ from the central position of a 2MASS XSC source even if it falls within its isophotes.

\subsubsection{WISE Extended Source Catalog (WXSC)}
The AllWISE photometric catalogues are optimized for the characterization of point-sources\footnote{AllWISE Explanatory Supplement: \href{https://wise2.ipac.caltech.edu/docs/release/allwise/expsup/sec2_2.html}{Cautionary Notes}}. For highly extended sources, source detection and extraction may not include all the extended components of the galaxy in a single source, thus leading to an underestimation of true brightness.

The WXSC \citep{Jarrett2019} provides full source characterization in all four WISE bands for the 100 largest galaxies (in angular extent) in the sky. WXSC uses new mosaics with native resolution allowing for precise measurements of both the target galaxy and its local environment. These mosaics are further resampled with 1" pixels, greatly enhancing analysis accuracy.  When a galaxy is present in WXSC, we use these magnitudes instead of those from AllWISE.

\subsection{WISE2MBH parent sample}\label{sec:xmatch}
All sources in the ETHER sample at $z\leq0.5$ were cross-matched with the AllWISE and WXSC catalogues with a cone search radius of 3\arcsec, preferring WXSC matches over AllWISE to consider better photometric values. After deleting duplicates, $95.2\%$ of the ETHER targets matched the two WISE catalogues. 

Given that the AllWISE catalog has a high density of sources (on average $\sim5$ sources per arcmin$^2$) and the ETHER sample at z $<$ 0.5 is also relatively large ($\sim$2.3 million galaxies, with a bias towards northern galaxies), it is important to address the probability of false matches in a $3\arcsec$ radius. We generate random samples of $10^5$ right ascension and declination coordinate pairs, divide them into subsamples in ($l$ between $\pm15$) and outside the Galactic plane, and test for random matches and possible contamination of W1 magnitudes due to confusion. When matching these random samples to the AllWISE catalog, we find a $4\%$ ($\sim5\%$) chance of matching a random position source out of the plane (in the plane) of the galaxy with an AllWISE source. The distributions of W1 magnitudes for real ETHER galaxies matched to WISE are consistently brighter than the magnitudes obtained in the random matches. The W1 magnitude distributions of both AllWISE sources matched to the random position catalogues ($\mu=17$ mag) and to the ETHER galaxies ($\mu=14$ mag), are significantly fainter than those of Galactic AGB stars in AllWISE \citep[$\mu=8$ mag;][]{Suh2021}. 

The differentiation between early-type galaxies and stars in AllWISE catalogues, in the absence of redshifts, remains a topic of ongoing discussion. Machine learning classification techniques have exhibited significant achievements in this area \citep[e.g.,][]{Kurcz2016}. In our case we do not expect to be affected by this problem for two reasons (a) ETHER has both precise (typically sub-arcsec accuracy) coordinates and includes only known extragalactic objects (when redshifts exist these are spectroscopic); (b) the low probability of random matches as explored in the previous paragraph. 

A summary of the resulting WISE2MBH parent sample is given in Table \ref{table:summary} where we list the total number of sources of each object type ($N$), and the total number of sources of the WISE2MBH parent sample and ETHER. The other columns represent a percentage of completeness.  
The distribution of sources in redshift by object type and \texttt{ex} can be seen in Fig. \ref{fig:hist_data_input}.

\section{K-corrections for WISE magnitudes and colors}\label{sec:kcorrections}

\citet{Jarrett2023} used spectral energy distribution (SED) templates of galaxies with different morphological types \citep[e.g.,][]{Brown2014} to calculate K-corrections for the W1 magnitude, and the W1$-$W2 and W2$-$W3 colors, for each available $T_\text{Type}$ over the range $z = 0-3$. These K-correction lookup tables are shown in Figs. 9 and 10 of \citetalias{Jarrett2023}, but only for $z\leq0.5$. 

Due to the uncertain morphological type and AGN contamination of objects classified as `RS', and given the tendency of QSOs to continue to reside in the AGN/QSO area of a WISE color-color diagram independent of redshift \citep[e.g.,][]{Mateos2012, Stern2012}, we do not apply K-corrections to the RS and QSO object types.  

\begin{figure}
    \centering
    \includegraphics[width=0.48\textwidth]{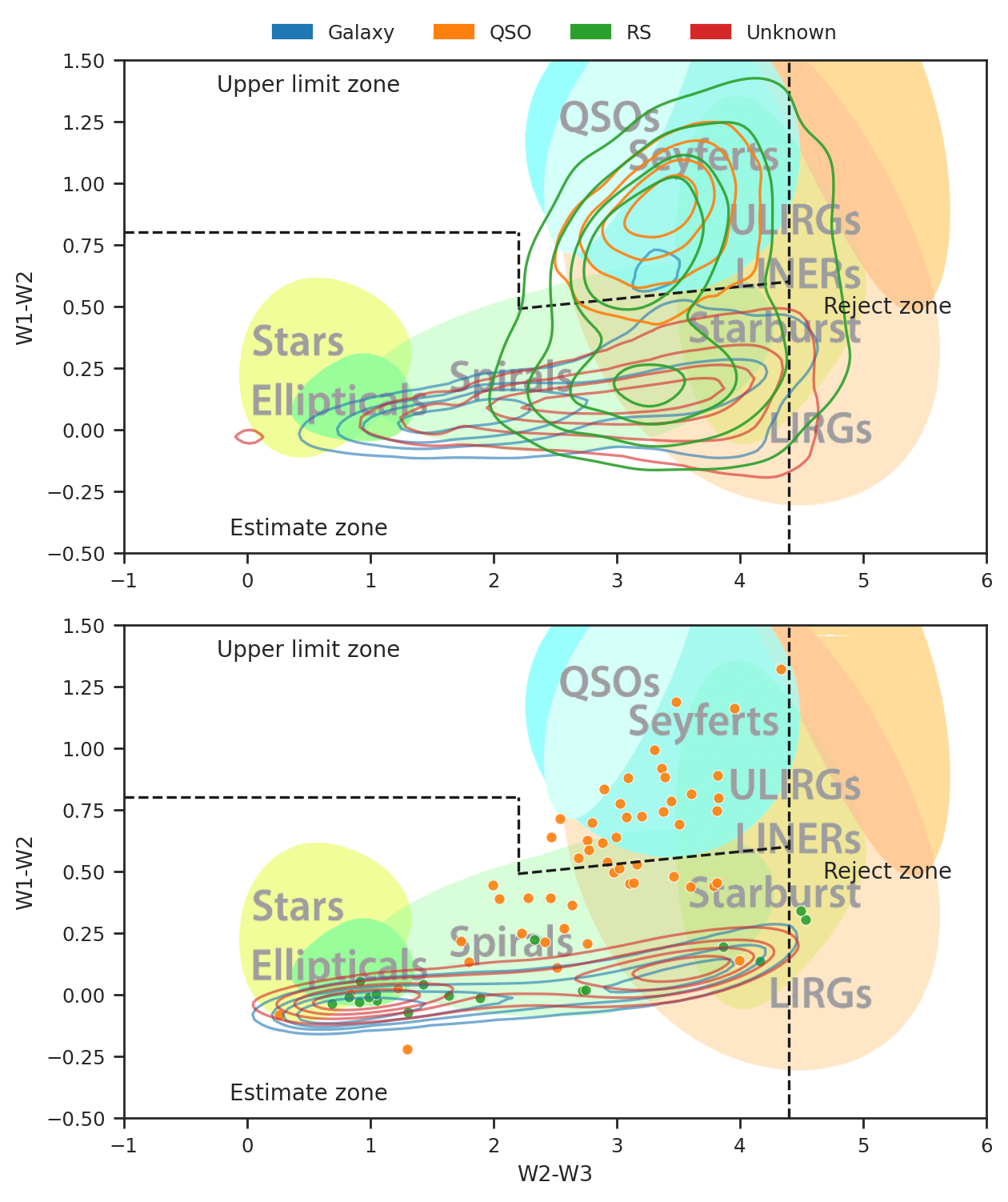}
    \caption{WISE color-color plot showing the location of our sample objects and defining the areas over which the stellar mass estimation is an upperlimit (Upper limit zone), is not estimated by the algorithm (Reject zone), or estimated by the algorithm (Estimate zone); these zones are separated by  black dashed lines. 
    Note that QSO and RS object types are considered as stellar mass upper limits independent of whether they fall in the "Upper limit" or "Estimate" zone. 
    For clarity, we plot separately the point sources (top panel) and extended sources (bottom panel; see Sect. \ref{sec:data}). 
    The background-filled colors and labels are from Fig. 12 of \citet{Wright2010}, and are shown for reference. Contours, in colors following the color legend on top of the figure, show the number density of object types in our sample. 
    In the bottom panel, the object types QSO and RS are shown as colored points (instead of contours) of the corresponding color.} 
    \label{fig:colorcolor}
\end{figure}

Of the remaining objects (i.e., galaxies or unknown) only those which reside in the `Estimate zone' of Fig. \ref{fig:colorcolor} (see Sect. \ref{sec:ident-sources}) are K-corrected. This selection is made to avoid a false K-correction for AGN-contaminated sources. 

Our algorithm applies the \citetalias{Jarrett2023} K-corrections for three  $T_\text{Type}$ ranges; ellipticals ($-$5 to $-$3), lenticulars (from $-$3 to 0) and spirals (from 0 to 8). The existing $T_\text{Type}$ of an individual object, if available, is used to select between these three lookup tables.
If no $T_\text{Type}$ is available, the lookup table to be used is decided as follows.
\citet[see their Fig. 5]{Mateos2012} use galaxy templates of various morphological types, and with varying amounts of AGN contamination, to derive the redshift evolution of a galaxy's WISE colors over the redshift range 0 and 2. Although the WISE colors vary with redshift (especially for late-type galaxies), they find that, in the absence of a large AGN contamination, WISE colors can distinguish between early-type and late-type galaxies over this full redshift range. 
We thus define the following cutoff in the observed W2$-$W3 color
to distinguish between elliptical and spiral galaxies (without AGN contamination) up to $z\sim1.5$, 
\begin{equation}
    \text{W2$-$W3}_\text{Limit} = -2.5\left(\log\left(\frac{F_\text{W3}}{F_\text{W2}}\right)-\log\left(\frac{f_\text{W3}}{f_\text{W2}}\right)\right)
    \label{eqn:wcolor}
\end{equation}
where $F_0$ and $f_0$ are the zero magnitude flux density and observed flux density in the respective band, respectively. 
We set the logarithmic ratio of observed flux densities in Eq. \ref{eqn:wcolor} to $-0.1$, since Fig. 5 of \citet{Mateos2012} shows that this value clearly separates elliptical and spiral galaxies over their full redshift range. This translates to a cutoff value of $\text{W2$-$W3}_\text{Limit} \sim 1.58$ to distinguish elliptical and spiral galaxies (values larger than this imply a spiral galaxy). Note that this cutoff is similar to the values previously derived by \citet{Jarrett2019}. The Lenticular lookup table is not used in this case, since lenticulars and spirals are not easily distinguishable in WISE color-color diagrams (see e.g., Sect. \ref{sec:est-ttype}).

These K-correction look up tables have been shown to be reliable for galaxies at $z\leq0.5$ \citep[e.g.,][]{Jarrett2023, Karademir2023}.

\section{Distinguishing Galaxies and powerful AGN/QSOs using WISE}\label{sec:ident-sources}
If an object type is not already available (which is the case for $76\%$ of the parent sample; Sect. \ref{sec:data}), the algorithm uses WISE color-color criteria to distinguish galaxies from powerful AGN/QSOs, and to determine whether the derived values are estimates or upper limits (due to contamination from an AGN). 

The algorithm is first tested on targets with an available object type. As mentioned in Sect. \ref{sec:kcorrections}, we use cutoffs in the observed W1$-$W2 and W2$-$W3 colors to identify objects that will be K-corrected. The same cutoffs are now applied to the K-corrected colors to determine if this target falls within the estimate, upper limit or rejection zones in the WISE color-color plot shown in Fig. \ref{fig:colorcolor}.
These objects then, respectively, follow the estimate, upper limit or reject paths of the algorithm shown in Fig. \ref{fig:flowchart} and described in Sect. \ref{sec:est-stellar-bulge}.

Objects with object type QSO and RS are likely to have significant AGN contamination in their W1 mag, which will likely increase their $M_*$ estimates, independent of their extension in WISE. For these objects, we therefore immediately consider the WISE2MBH $M_*$ values as upperlimits.
The horizontal dashed line in Fig. \ref{fig:colorcolor}, which separates the upper limit zone from the estimate zone, is a combination of the widely used $\text{W1$-$W2} = 0.8$ limit to separate AGN/QSOs from galaxies \citep[e.g.,][]{Stern2012, Michalik2016, Guo2018}, together with a wedge region between W2$-$W3 $\sim$2.2 -- 4.4 motivated by previously defined AGN/QSOs regions \citep{Jarrett2011, Hviding2022}. 
We slightly modified the wedge region defined by \cite{Jarrett2011} by optimizing the WISE color classification of known QSOs in our sample. The wedge region we chose is defined as follows
\begin{equation}
    2.2<\text{W2$-$W3}<4.4 \wedge \text{W1$-$W2}>(0.05[\text{W2$-$W3}]+0.38). 
\end{equation}
which now fully extends to the blue-side of the diagram thanks to the $\text{W1$-$W2} = 0.8$ cutoff.
With this overall cut, almost $95\%$ of the known QSOs in our sample are classified as QSOs by this color criterion. 

The vertical limit at $\text{W2$-$W3}=4.4$ in Fig. \ref{fig:colorcolor}, which separates the Reject zone from the other two zones, is also set by us. At $\text{W2$-$W3} \geq 4.4$, the WISE2MBH $T_\text{Type}$ (see Sect. \ref{sec:est-ttype}) is 8 or larger, and the estimated $B/T$ would be very low (Sect. \ref{sec:est-bt}).
While this would be correct for, e.g., irregular galaxies, it is incorrect for, e.g., extremely dusty, hybrid-starburst-AGN galaxies \citep{Tsai2015} or newborn AGN (\textcolor{blue}{Arevalo et al. in prep})\footnote{P. Arevalo showed in XVII LARIM on December 2023, the evolution in W1$-$W2 color for a typical star-forming non-active galaxy to AGN-like in a time-span of years, together with spectroscopic data from SDSS and SOAR that support the idea of new AGN activity in \href{https://alerce.online/object/ZTF20aaglfpy}{ZTF20aaglfpy}, which was also classified as a type I AGN by the ALeRCE light curve classifier.}.

The WISE color-color distributions of different object types (including unknown types) in our sample are shown in the top (WISE point sources) and bottom (WISE extended sources) panels of Fig. \ref{fig:colorcolor}. 

The top panel of Fig. \ref{fig:colorcolor} shows that most QSOs and RS reside at high W1$-$W2 colors, in an area populated by a variety of AGN and also ULIRGs, making it a challenge to correctly identify them using only WISE data. On the other hand, known galaxies, even if point-like for WISE, show clear overdensities in the region where elliptical and spiral galaxies are expected to reside.

For the bottom panel of Fig. \ref{fig:colorcolor} (WISE extended sources), all ($\sim99\%$) known galaxies lie in the Estimate zone, at WISE colors expected of elliptical and spiral galaxies, and away from the regions of starbursts, LINERs and (U)LIRGs. Only a few QSOs and RS are present in this figure: the QSOs (orange dots) do not clump in the expected QSO area, but are instead distributed over a large range of colors, overlapping with regions of galaxies, ULIRGs, and Seyferts; RS (green dots) are predominantly situated in the galaxies area, with a few in the LINER area. This tendency for both QSOs and RS to not only reside in the expected wedge region described above is not unexpected; since these extended sources are most likely weaker AGNs whose emission is not sufficient to change the galaxy color so as to be classified as a QSO. This was shown in \cite{Mateos2012} who tested different percentages of AGN activity, and showed that only powerful AGNs ($80\%$ fraction) up to $z=2$ reside in an area similar to the wedge shaped region defined here or in previous studies.

The bulk of sources with unknown type ($76\%$ of the parent sample; red contours in both panels) sit in the Estimate zone, mostly following the expected distribution of galaxies. The clear separations seen for known object types give us confidence that the classification of these unknown types as galaxies is reliable.

\section{Using W2-W3 color to estimate T-type}\label{sec:est-ttype}
While $T_\text{Type}$ is available for some ($\sim22\%$) sources in the WISE2MBH parent sample, a reliable estimation of this is required for most of the sample. To obtain these $T_\text{Type}$, we exploit the fact that the W2$-$W3 color shows clearly separated regions where elliptical and spiral galaxies reside \citep[e.g.,][]{Wright2010,Jarrett2019,Cluver2020}. Although these regions partially overlap with other classifications based on star formation activity, the trend is sufficient to estimate the morphology of the galaxy, i.e., $T_\text{Type}$.
\begin{figure}
    \centering
    \includegraphics[width=0.48\textwidth]{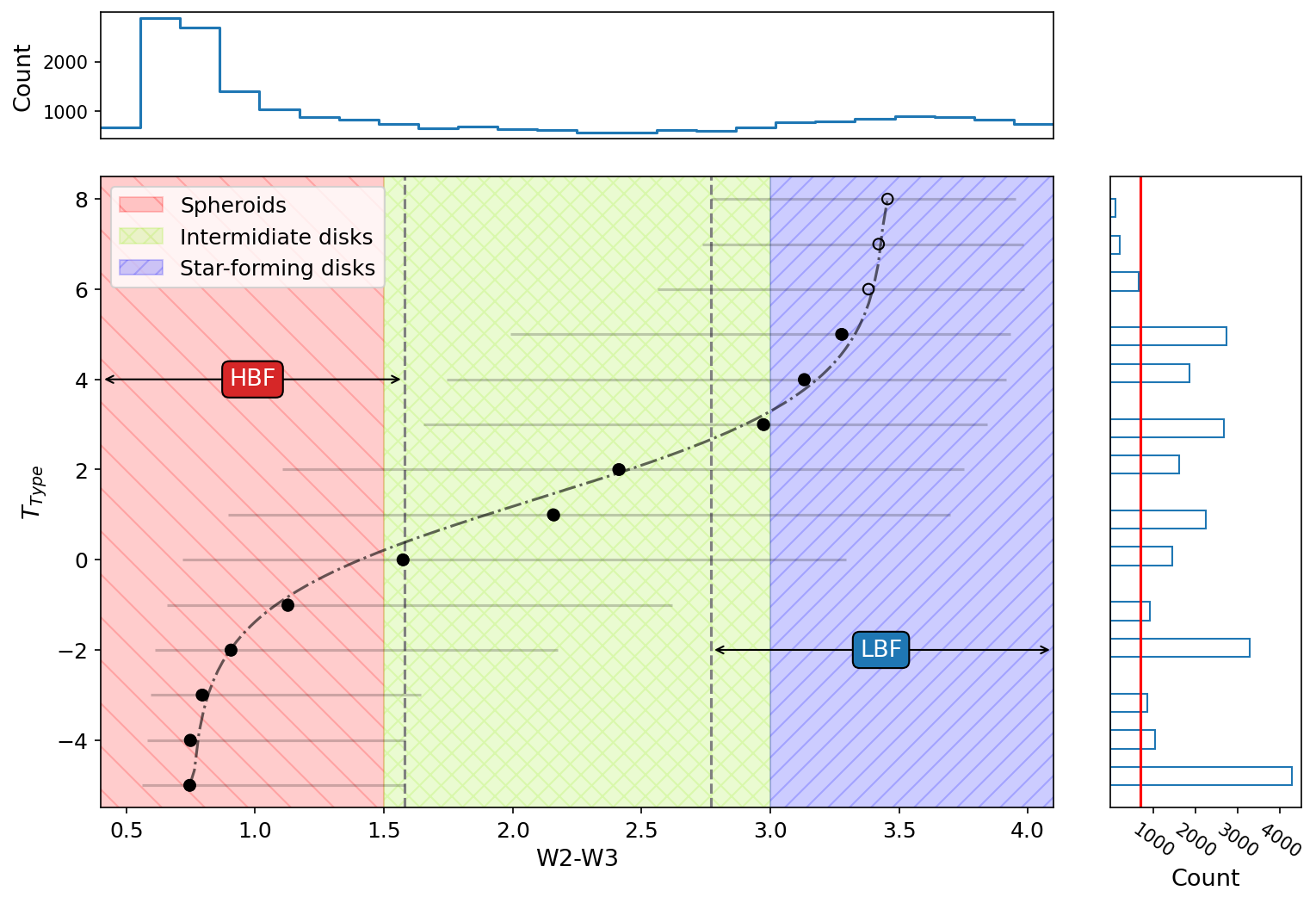}
    \caption{$T_\text{Type}$ as a function of W2$-$W3 color for galaxies in the 2MRS sample. For each $T_\text{Type}$ bin we plot the median value (black circle) and one sigma dispersion (horizontal bar) of the W2$-$W3 colors of galaxies in the bin. 
    Distributions of W2$-$W3 and $T_\text{Type}$ are shown in the panels at the top and right of the figure, respectively.
    The red line in the right panel marks the threshold number of galaxies in a bin in order for that bin's median to be used for the fit (filled black circles in the main panel). 
    The black dot-dashed line shows the best fit logit function to the filled black circles: this is used for the W2$-$W3 to $T_\text{Type}$ conversion when $T_\text{Type}$ is previously unknown. The estimated  $T_\text{Type}$ is limited to the range $-$5 and 8; when a galaxy's W2-W3 color lies beyond the range of the logit function shown, the  $T_\text{Type}$ is clipped at these values. 
    The colored areas distinguishing morphologies listed in the inset are from \protect\cite{Jarrett2019}.
    Given the similarity of the color dispersions in the three bins at each extreme end of the $x$-axis, we define two vertical dashed lines which delineate galaxies we refer to as high bulge fractions (HBF; bulge fractions between 0.4 and 1) and low bulge fractions (LBF; bulge fractions between 0.1 and 0.3). 
    }
    \label{fig:logit}
\end{figure}
\begin{figure}
    \centering
    \includegraphics[width=0.48\textwidth]{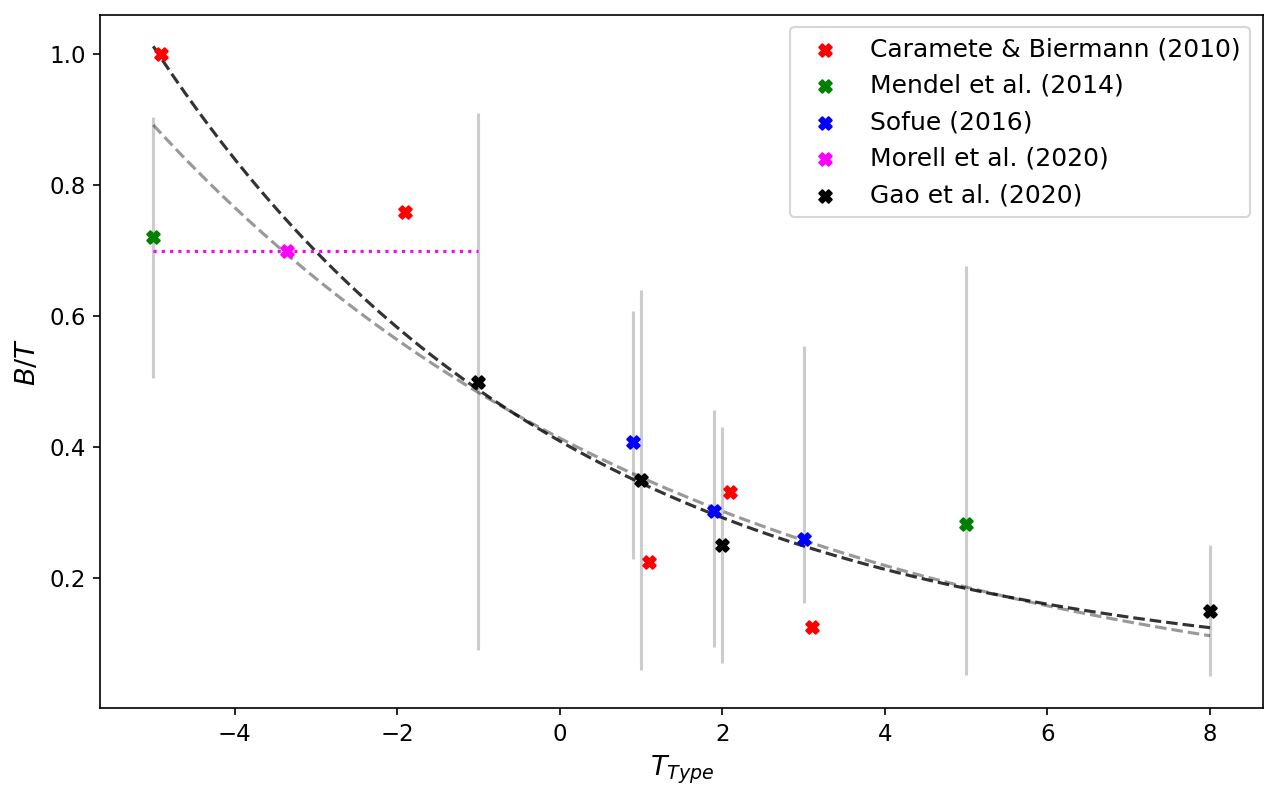}
    \caption{Distributions of the bulge-to-total mass ratio ($B/T$) as a function of $T_\text{Type}$ for different literature samples. A decreasing exponential is fitted to the data points: the gray dashed curve is the original fit and the black dashed curve is the fit when one fixes $B/T=1$ for $T_\text{Type}=-$5. For clarity, small shifts on the x-axis are used to avoid overlapping symbols and error bars. Data points and the horizontal pink dashed line are from \citet{Caramete2010, Mendel2014,Sofue2016,Morell2020,Gao2020} following the colors listed in the inset.}
\label{fig:t_vs_bt}
\end{figure}

Our conversion of W2$-$W3 color to $T_\text{Type}$ is trained using $\sim$18,000 galaxies from the 2MRS catalog \citep{Huchra2012} for which manually classified $T_\text{Type}$ are available. The median W2$-$W3 colors of 2MRS galaxies in each $T_\text{Type}$ bin between $-5$ and $8$ (open and filled black circles in the main panel of Fig. \ref{fig:logit}) show a clear S-shape curve. 
The number of galaxies in each $T_\text{Type}$ bin is shown in the right histogram of Fig. \ref{fig:logit}. 
Given the S-shape, i.e., the lack of differentiation in the $x$-axis for the three to four bins at each extreme end of W2-W3 colors, a sufficiently large number of objects per bin is required for a clear result. We therefore use statistical power analysis to define the required sample size threshold; details of this analysis can be found in Appendix \ref{sec:appen1}. For this power analysis, a power $P=0.8$ and a significance threshold $\alpha=0.05$ were assigned. The effect size ($E_s$) was calculated for each set of two consecutive bins in W2$-$W3 (in order of increasing $T_\text{Type}$), and the resulting median $E_s$ (0.15) implies that the sample size per bin must be $N\gtrsim700$. Therefore, all bins with a sample size greater than this $N$ were accepted (filled black circles in Fig. \ref{fig:logit}). 

A logit function was fit to these accepted median values, providing us with our W2$-$W3 to $T_\text{Type}$ conversion. Since the logit function's domain goes from 0 to 1, the W2$-$W3 color is shifted and normalized ($\text{W2$-$W3}_\text{SN}$) before fitting: 
\begin{equation}
\begin{split}
    &\text{W2$-$W3}_\text{SN} = \frac{\text{W2$-$W3} - 0.75}{2.71}\\
    &T_\text{Type} = (1.21\pm0.01)\text{ logit} \left(\text{W2$-$W3}_\text{SN}\right) + (1.36\pm0.02)
\end{split}
\label{eqn:color-ttype}
\end{equation}
this logit function is shown as a dashed-dotted line in Fig. \ref{fig:logit}. Despite leaving out three late-type $T_\text{Type}$ bins from the fit ($T_\text{Type}$ = 6, 7 and 8), the logit function fits almost perfectly to all medians.

The morphological limits of Fig. 10 of \cite{Jarrett2019} are presented for comparison as colored regions in the figure. While the overall fit is S-shaped, we find two clearly separated regions in the graph: the high bulge fraction (HBF) region, which is delimited by the 84$^\text{th}$ percentile values of the bins centered on $T_\text{Type}$ = $-5$ to $-3$, and the low bulge fraction (LBF) region, which is delimited by the 16$^\text{th}$ percentile values of the bins centered on $T_\text{Type}$ = $7$ and $8$. Our HBF region limit ($\text{W2$-$W3}=1.58$; vertical dashed line in the figure) is similar to the cutoff used by \cite{Jarrett2019} to distinguish between spheroids and intermediate disks (the division between pink and green regions in the figure), and also similar to the value of $\text{W2$-$W3}_\text{Limit}$ which we use to classify galaxies with unknown morphological type into elliptical and spiral galaxies in order to select the K-correction lookup table to be used (see Eq. \ref{eqn:wcolor} and Sect. \ref{sec:kcorrections}). 

The logit function in Eq. \ref{eqn:color-ttype} is used whenever a source does not have an available $T_\text{Type}$ or if the available $T_\text{Type}$ comes from a binary classification \citep[e.g.,][]{Dobrycheva2013}.
The WISE2MBH algorithm uses $T_\text{Type}$ in the range $-5$ to $8$. Available $T_\text{Type}$ values outside this range are clipped to the closest limit value, in case these really define a morphology, e.g., $-9$ is often used to define a QSO (ZCAT convention), so those values are not clipped.
If a measured W2$-$W3 color is outside the range of the logit function presented here, the estimated $T_\text{Type}$ is also clipped to the closest limit value $T_\text{Type}$. 
This is most relevant for elliptical galaxies, whose W2$-$W3 colors are often less than $0.7$, which the algorithm converts to $T_\text{Type}=-5$.

This $T_\text{Type}$ estimator in WISE2MBH is an auxiliary function: it prefers an input value but will provide a WISE-based value when necessary. The S-shape curve makes the classifications at extreme $T_\text{Types}$ very uncertain at the moment of distinguishing between consecutive $T_\text{Types}$, but the distinction between bulge-dominated and disk-dominated galaxies is clear.

\begin{figure*}
    \centering
    \includegraphics[width=\textwidth]{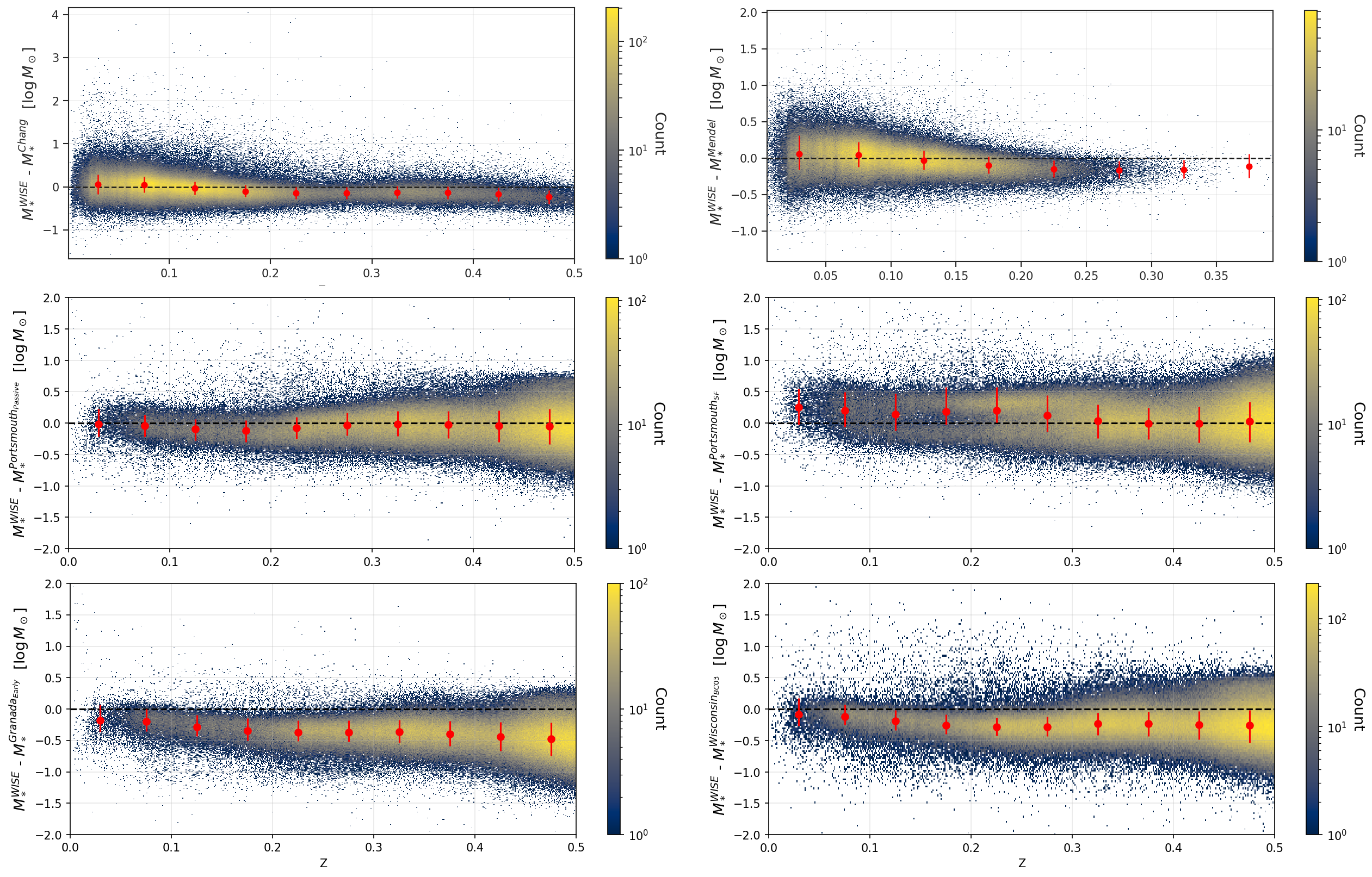}
    \caption{WISE2MBH $M_*$ compared to the low redshift ($z\leq0.5$) galaxy samples of \citet[Top left]{Chang2015} and \citet[Top right]{Mendel2014} and the SDSS Value Added catalogues of the Portsmouth \citep[][Middle]{Maraston2013}, Granada \citep[][Bottom right]{Chen2012} and Wisconsin \citep[][Bottom left]{Montero2016} groups. The black dashed line shows the line of equality and red dots with error bars represent the median and 1$\sigma$ dispersion of the difference between the respective masses for bins of 0.05 in redshift. Colors represent the counts following the color bar to the right of each panel.}
    \label{fig:stellar_comp}
\end{figure*}

\section{Bulge-to-Total ratio from T-Type}\label{sec:est-bt} 
The morphological type of a galaxy within the Hubble sequence has been shown to be a good proxy of $B/T$. This inverse behavior \citep[recently discussed in][]{Quilley2022,Quilley2023} shows that early-type galaxies tend to be almost pure bulges ($B/T\sim1$), while very late-type galaxies and irregulars tend to have small to null bulge fractions ($B/T\sim0.01$).
This inverse behavior also supports the posited scenarios of galaxy bulge growth via mergers: late-type galaxies merge consecutively until lenticular, elliptical, and finally, the brightest cluster galaxies (BCGs) are formed \citep{Edward2020}, leading not only to the formation of the most massive galaxies \citep{Bluck2014}, but also to the most massive SMBHs \citep{Mezcua2018}.

By nature, elliptical galaxies are expected to have quenched their star formation, leading to a decrease in their SFR and specific SFR (sSFR), despite environmental effects \citep[e.g.,][]{Casado2015}. Recently, \cite{Ge2018} showed that galaxies with lower sSFR tend to be more massive and have higher $B/T$ ($\geq0.7$) compared to galaxies with higher sSFR, and also found a trend with galaxy age where the oldest galaxies have higher $B/T$. \cite{Morell2020} showed similar results, showing that their passive galaxy sample (made up of 70\% ellipticals and 15\% lenticulars) is the one with higher $B/T$ ($\sim0.7$).

Massive elliptical galaxies are the most relevant sources for future EHT observations, and while varying their estimated $B/T$ between 0.8--1 does not considerably affect the final $M_\text{Bulge}$ estimate, a misclassification of $T_\text{Type}$ could result in the incorrect use of a spiral-like $B/T$ ($\leq0.2$), leading to an incorrectly low $M_\text{Bulge}$, thus $M_\text{BH}$ \citep{Bluck2014}. 
Despite many references pointing to a $B/T\leq0.8$ for early-type galaxies \citep[e.g.,][]{Ge2018, Morell2020}, we will impose a limit of $B/T=1$ for $T_\text{Type}=-5$ \citep[e.g.,][]{Caramete2010,Quilley2023}, with an exponential decrease with $T_\text{Type}$ as shown in Fig. \ref{fig:t_vs_bt}.
We note that \cite{Caramete2010} derived correction factors to convert near-IR luminosities to $M_\text{BH}$ for galaxies of a variety of $T_\text{Type}$ in the nearby universe ($z\leq0.025$) with similar impositions for elliptical galaxies; our $B/T$ ratios are analogous to their conversion factors but calibrated with a larger sample.   

We calibrate the $B/T$ ratio as a function of $T_\text{Type}$ using several samples from the literature. 
\cite{Mendel2014} provide total, bulge, and disk masses, for a large sample of SDSS galaxies. Their values, combined with $T_\text{Type}$ from ETHER, give us distributions of $B/T$ over a wide range of $T_\text{Type}$, although only $-5$ and $5$ have enough statistics to be considered robust. 
The ETHER $T_\text{Type}$ unfortunately comes primarily from the binary classification of \cite{Dobrycheva2013}, so we expect that the relationship between $T_\text{Type}$ and $B/T$ is biased, i.e., underestimated for ellipticals and overestimated for spirals. \cite{Sofue2016} provide both bulge and disk masses for a small sample of nearby galaxies ($z\leq0.03$) for which we obtained $T_\text{Type}$ from NED. \cite{Morell2020} provide an average value of $B/T$ for a sample of passive galaxies, which are composed of specific fractions of ellipticals, lenticulars and spirals, mostly dominated by the first two. The $T_\text{Type}$ value we used in this case is weighted by these fractions. From \cite{Gao2020} we take values of $B/T$ only for $T_\text{Type}$ equal to $-1$, $1$ and $2$ (S0, Sa and Sab) and consider $T_\text{Type}$ equal to $8$ (Sdm) as a limit to secure reasonable bulge masses even for very late-type galaxies. 

A plot of $T_\text{Type}$ as a function of $B/T$ for all these samples is shown in Fig. \ref{fig:t_vs_bt}. A direct fit to all these points is shown with the dashed gray line in the figure. When we force a value of $B/T = 1$ for $T_\text{Type}$ = $-5$, our final fit is functionally similar to that of \citet[Fig. 1]{Caramete2010}, but now for significantly larger and more recent datasets. The final fit is,
\begin{equation}
    B/T = 0.05 + 0.36\left(7.72^{-0.1\cdot T_\text{Type}}\right)
\end{equation}
which results in a $13\%$ increase in $B/T$ at $T_\text{Type}$ = $-5$ compared to the direct fit to all datapoints, while for $T_\text{Type}$ $\geq-1$ the difference is negligible.

\begin{figure}
    \centering
    \includegraphics[width=0.48\textwidth]{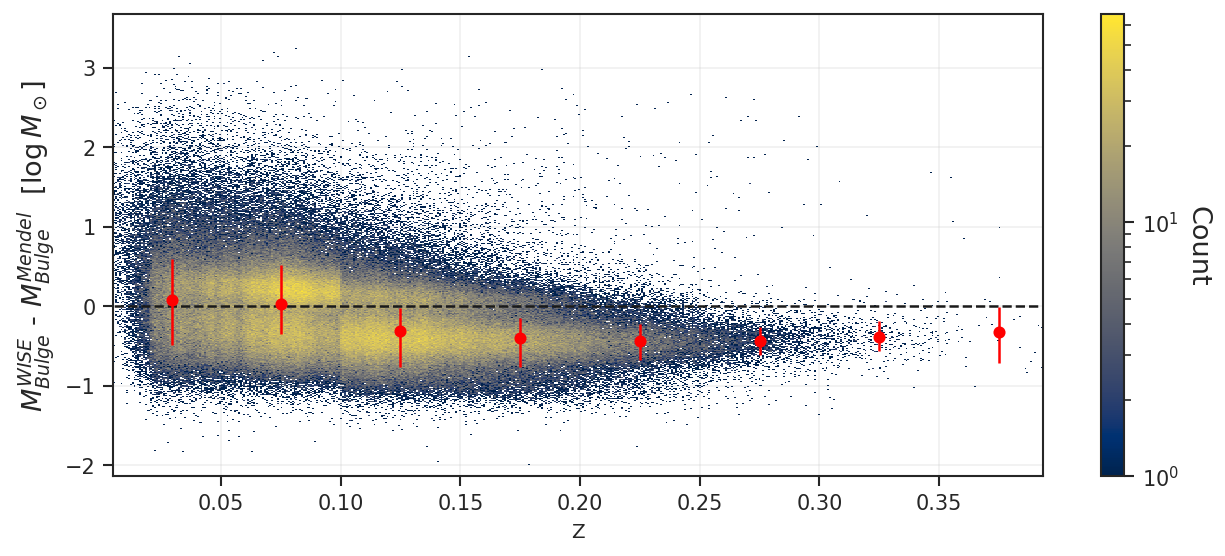}
    \caption{Same as Fig. \ref{fig:stellar_comp}, but now comparing $M_\text{Bulge}$ from  WISE with its equivalent value in the low redshift ($z\leq0.4$) sample of \citet{Mendel2014}.}
    \label{fig:bulge_comp}
\end{figure}

\section{Stellar and Bulge mass from WISE photometry}\label{sec:est-stellar-bulge}
The process of converting WISE photometry to stellar mass is described in \citetalias{Cluver2014}, who take advantage of the fact that W1 is an exceptional tracer of the bulk of stellar population in galaxies and that the W1$-$W2 color can constrain the M/L ratio. For their (and our) calculation, the W1 absolute magnitude of the Sun is taken from \cite{Willmer2018}. \citetalias{Jarrett2023} have presented an updated $M_*$ estimator valid across a larger redshift range, making use of multi-color criteria and K-corrections. We use the K-corrections of the latter (see Section \ref{sec:kcorrections}) for W1 and W1$-$W2 colors together with the stellar mass estimator of \citetalias{Cluver2014} to derive the total stellar mass ($M_*$).

To avoid excessive (and likely erroneous) M/L values estimated from W1$-$W2 colors, our algorithm limits the input W1$-$W2 values to the range $-0.2$ to $0.6$ (corresponding to high and low M/L). Any source with a W1$-$W2 color outside this range is clipped to the nearest limit M/L, i.e., the distribution of W1$-$W2 (generated by the random normal samples of W1 and W2, see Fig. \ref{fig:flowchart} and Sect. \ref{sec:algorithm}) is shifted until the median reaches the closest limit value.

Once $M_*$ and its errors are calculated following the process outlined above, the value is stored unless the estimated mass is $\log M_*\leq6.5$ or $\log M_*\geq13$. 
This range is more strictly constrained at the low mass end than other catalogues, and more lax at the high mass end \citep[e.g.,][]{Dimauro2018, Durbala2020}. The flexibility at the high mass end is in order to not lose extremely rare extreme SMBHs, often called ultra massive black holes \citep[UMBHs, e.g,][]{Runge2021} and SMBH upper-limits, e.g., QSO with high M/L, which produce very high $M_\text{BH}$ estimates, but which are flagged as upper limits. 
The lower limit value is extracted directly from \citetalias{Jarrett2023}.

The WISE2MBH $M_*$ estimates derived here are compared to two low redshift control samples from \citet[$z\leq0.4$]{Mendel2014} and \citet[$z\leq0.5$]{Chang2015} in Fig. \ref{fig:stellar_comp} and to a group of SDSS samples \citep{Chen2012,Maraston2013,Montero2016}, all for $z\leq0.5$. For the samples of \citeauthor{Mendel2014} and \citeauthor{Chang2015}, the agreement is relatively good, with a scatter of $\sim0.2$ dex in both cases. There is a slight tendency for WISE2MBH $M_*$ to be underestimated with increasing redshift compared to some of the SDSS samples, No strong evidence was found to suggest that the estimates from SDSS  are superior to those from WISE2MBH; correlations and scatters between these samples did not demonstrate a preference. Appendix \ref{sec:appen3} contains a more detailed discussion of this WISE to SDSS comparison. 

The $M_\text{Bulge}$ estimations are obtained by combining $B/T$ with $M_*$. Effectively, the estimated $B/T$ allows the estimation of both $M_\text{Bulge}$ and $M_\text{Disk}$. In the WISE2MBH final catalog, we provide only $M_*$ and $B/T$ for simplicity. Figure \ref{fig:bulge_comp} presents a comparison of WISE2MBH bulge masses with those derived by \cite{Mendel2014} in a low redshift sample: once again, the agreement is relatively good, but now shows an increased scatter. This can be explained by our assumption of a simple evolution of $B/T$ with $T_\text{Type}$, which does not consider other important factors in galaxy evolution such as gas availability, molecular gas content, size distribution, stellar age, and the impact of bars and bulges  \citep[e.g.,][]{Laurikainen2007,Fisher2011,Koyama2019}.

\begin{figure}
    \centering
    \includegraphics[width=0.48\textwidth]{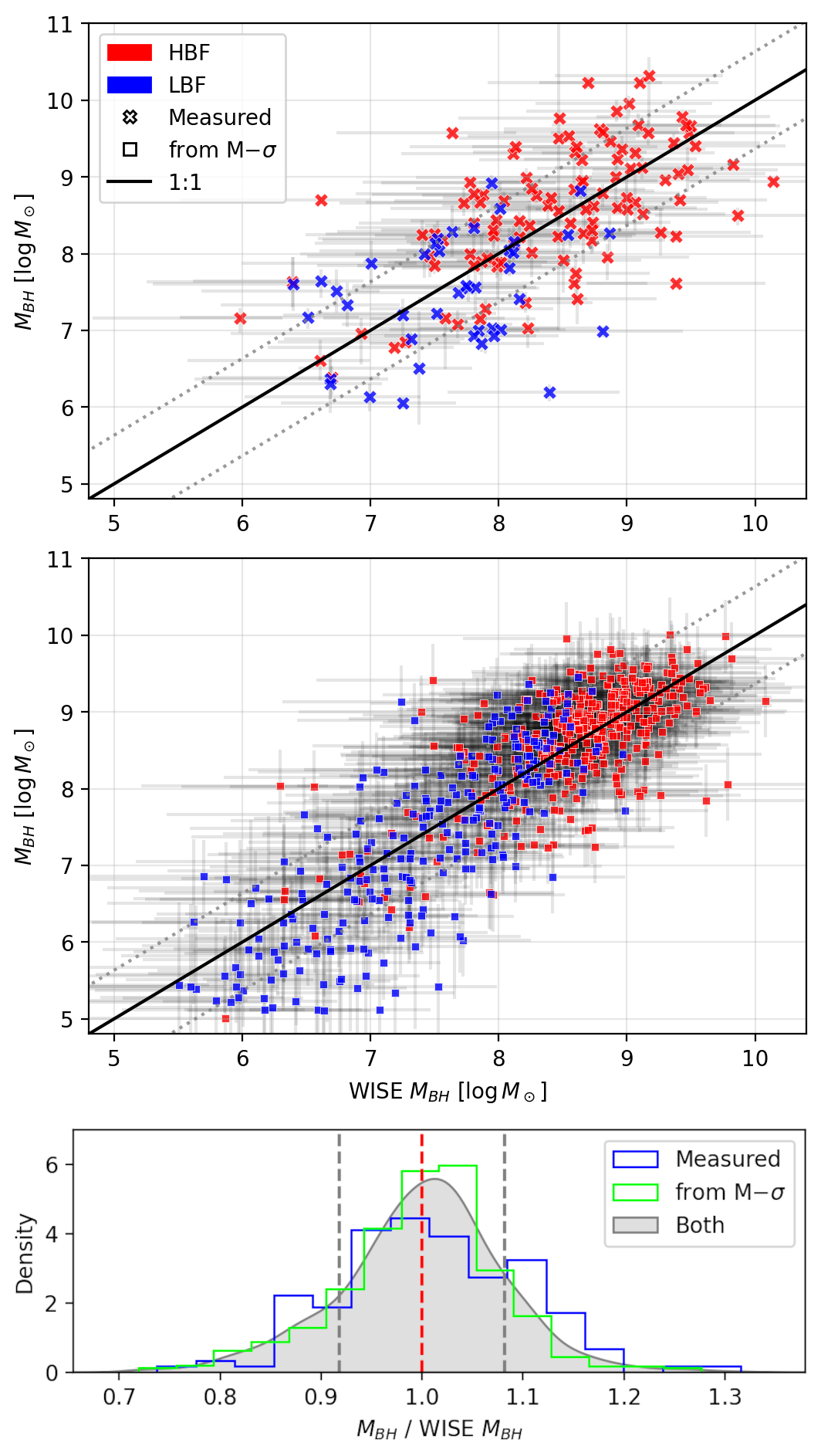}
    \caption{A comparison of measured black hole masses (Top, crosses) and highly reliable black hole mass estimates (Middle, boxes) vs. WISE2MBH $M_\text{BH}$ values in the WISE2MBH final sample. Each data point is marked with its $1\sigma$ error bars. Subsamples of HBF and LBF galaxies are distinguished by color following the inset. Gray dotted lines are the RMSE scatter bands. The bottom panel shows the distribution of the mass ratios for subsamples of measured and estimates and a KDE for the complete distribution; the mean ratio and 1 $\sigma$ dispersion ($1.00\pm0.08$) of the latter are shown with red and gray dashed lines.} 
    \label{fig:mbh_comp}
\end{figure}

\begin{figure}
    \centering    
    \includegraphics[width=0.48\textwidth]{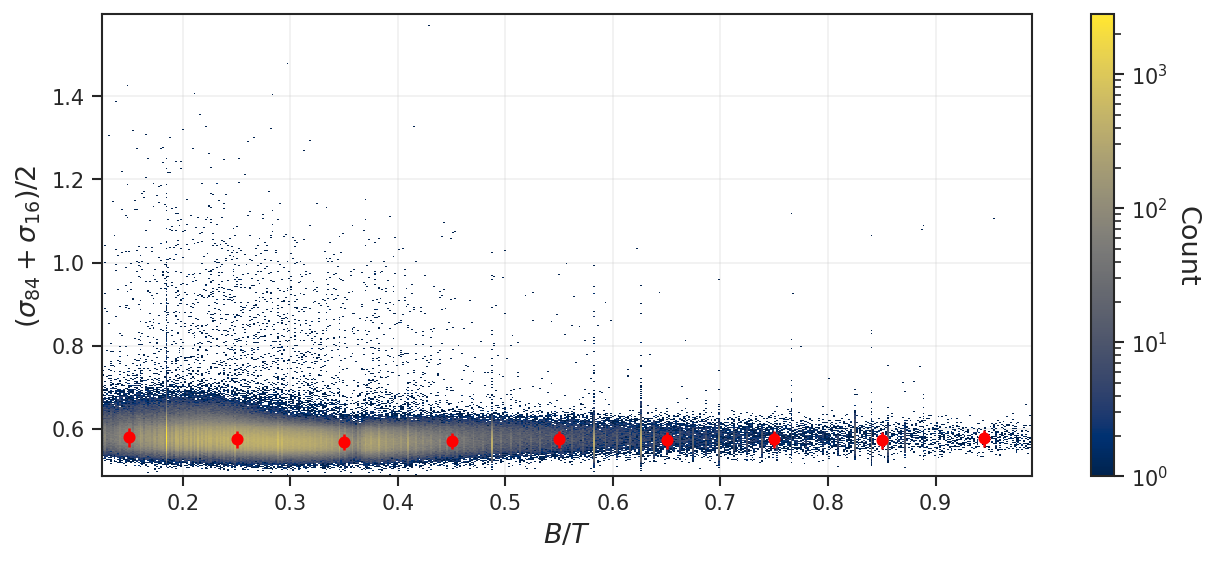}
    \caption{Mean $M_\text{BH}$ errors as a function of $B/T$. In LBF, there are few cases with larger error compared to the mean errors for both HBF and LBF. This comes directly from the detection of small sources that have worst quality than the bulk of sources detected by WISE in the AllWISE catalog, producing greater errors due to error propagation.
    } 
    \label{fig:err_per_bf}
\end{figure}

\begin{table}
\centering
\caption{Control sample of $M_\text{BH}$ measurements and reliable M$-\sigma$ estimations.}
\resizebox{\columnwidth}{!}{%
\begin{tabular}{@{}cccr@{}}
\toprule
$N$     & $M_\text{BH}$ / WISE $M_\text{BH}$    & Method           & Publication  \\ \midrule
46      & $1.01\pm0.08$ & \begin{tabular}[c]{@{}c@{}}Stellar / Gas dynamics\\ Megamaser\\ RM\end{tabular} & \cite{Bosch2016}  \\
13      & $1.03\pm0.10$ & Stellar dynamics & \cite{Saglia2016}    \\
6       & $0.97\pm0.05$ & Stellar dynamics & \cite{Thater2019}    \\
71      & $1.00\pm0.10$ & \begin{tabular}[c]{@{}c@{}}Stellar / Gas dynamics\\ Megamaser\end{tabular}      & \cite{Gultekin2019}   \\
16      & $1.06\pm0.08$                         & Gas dynamics     & ETHER compilation$^a$   \\\midrule
647     & $1.00\pm0.08$ & M$-\sigma$          & \cite{Bosch2015}    \\ \midrule
799     & $1.00\pm0.08$ &                  &                          \\ \bottomrule 
\end{tabular}%
}
\label{table:control_sample}
\flushleft\footnotesize{\textit{Notes:} $^a$From the ETHER sample \citnp{Nagar et al.}; see \citetalias{Ramakrishnan2023} for a description of the compilations.}
\end{table}
\section{Black Hole Mass from WISE photometry}\label{sec:est-bh}

The value of $M_\text{Bulge}$ calculated in the previous section is used to derive a first estimate of $M_\text{BH}$ using the $M_\text{BH}-M_\text{Bulge}$ relationship of \cite{Schutte2019}. 
These first $M_\text{BH}$ estimates were compared with a control sample presented in Table \ref{table:control_sample}. The control sample consists of $152$ galaxies with directly measured $M_\text{BH}$ from different methods \citep[e.g.,][]{Saglia2016,Bosch2016} and $647$ galaxies for which high-quality stellar velocity dispersions ($\sigma$) were available, obtained via observations with the Hobby-Eberly Telescope \citep[HET; ][]{Bosch2015}. Using these values of $\sigma$ allowed us to accurately estimate $M_\text{BH}$ using the M-$\sigma$ relationship of \cite{Saglia2016}.

The control sample was selected according to the following criteria: (a) the value of $M_\text{BH}$ is flagged as a measurement or high-quality estimate from $\sigma$, (b) the $M_\text{BH}$ estimate from WISE2MBH is not an upper limit, (c) the source has an \texttt{ex} flag equal to 5 and (d) the control sample source must have $\log M_\text{BH} \leq 10.32$. While points (a) and (b) are self-explanatory, (c) is required to consider only completely extended sources, and (d) is necessary since \cite{Bosch2015} contains a few very large  $\sigma$ values. To avoid those, we consider the maximum value from \cite{Bosch2016}, who used the same observations from HET as part of the study to measure $M_\text{BH}$, as a limit for our control sample. The control sample covers a mass range from $\log M_\text{BH}$ of $\sim$6 to $\sim$10, this being almost the complete mass range for SMBH. The heterogeneity of this control sample is discussed in Sect. \ref{secc:hetero}.

Linear regression done to the control sample $M_\text{BH}$ (dependent variable) and WISE2MBH $M_\text{BH}$ estimates (independent variable) revealed a slope of 0.9 and an intercept of 0.98. A t-test was then used to check the statistical significance of these results compared to a linear regression close to the equality line (expected slope of 1 and intercept of 0) with a p-value of $p=0.05$. The results showed that the slope did not differ significantly from the equality line, while the intercept was significantly different from zero, suggesting the need for a compensation factor. These findings demonstrate the presence of subtle, yet systematic, discrepancies between the WISE2MBH $M_\text{BH}$ estimates (when the \citet{Schutte2019} scaling is used) and the control sample values. To address these systematic offsets empirically, a compensation factor ($C_f$) is defined as follows,
\begin{equation}
    C_f = -0.104\log M_\text{BH} + 0.98 
\end{equation}
and added to the estimate. After this empirical correction, a Spearman score of 0.78 and a root mean squared error (RMSE) of 0.63 dex (see Fig. \ref{fig:mbh_comp}) were calculated for the set of compensated estimates and control sample. Since the majority of the control sample are HBF, the small offset of LBF shown at lower $M_\text{BH}$ ranges does not affect the overall comparison, but may be interpreted as the need for a specific $C_f$ for LBF sources or a misbehavior of previous steps for these types of galaxies, e.g. underestimation of $B/T$ or $M_*$.

The $M_\text{BH}-M_\text{Bulge}$ scaling relation of \citeauthor{Schutte2019} can be combined with our $C_f$, to obtain the following relationship, which is effectively that used in WISE2MBH:
\begin{equation}
    \log M_\text{BH} = 1.12\log\left(\frac{ M_\text{Bulge}}{10^{11}}\right) + 8.84 
\label{eqn:finalmbh}
\end{equation}
Finally, in case the algorithm estimates a $\log M_\text{BH}\leq5$, the source is dropped. 
While the $M_\text{BH}-M_\text{Bulge}$ relation of \cite{Schutte2019} can reach $\log M_\text{BH}\leq5$, such black holes go down to the limits of intermediate mass black holes (IMBH). IMBHs and their host populations are an active topic of research \citep[for a review, see][]{Greene2020}, and these populations do not necessarily follow the several scaling relations used in WISE2MBH. This also means that estimates slightly higher than $M_\text{BH}=5$  should be treated with caution. 

\section{Algorithm}\label{sec:algorithm}
The WISE2MBH algorithm was conceived with the purpose of addressing the lack of $M_\text{BH}$ estimates for more than $80\%$ of the ETHER sample (see Sect. \ref{sec:relevance-ngeht}). 
Nevertheless, it provides a simple and uniform tool with wide-ranging applications in studies of morphology and galaxy and black hole evolution. It is useful for both generation of large samples from existing data, and of sub-samples for future observations and monitoring with observational facilities. 

This main steps of the algorithm are summarized in Fig. \ref{fig:flowchart}. In summary, the process is as follows:
\begin{itemize}
    \item W1, W2 and W3 magnitudes and a spectroscopic redshift are used to calculate the K-corrected W1 absolute magnitude, W1$-$W2 and W2$-$W3 colors with the use of lookup tables from \citetalias{Jarrett2023} (see Sect. \ref{sec:kcorrections}).
    \item The K-corrected W1 absolute magnitude and the W1$-$W2 color are used to estimate $M_*$ of the source with the process described in \citetalias{Cluver2014}. For that calculation, the W1 absolute magnitude of the Sun is taken from \cite{Willmer2018} (see Sect. \ref{sec:est-stellar-bulge}).
    \item The source is placed into the estimate or upper limit zone, making use of several inputs (see Sect. \ref{sec:ident-sources}).
    \begin{itemize}
        \item If a source is placed in the estimate zone and its $T_\text{Type}$ is available, it continues in the algorithm.
        \item If a source is placed in the estimate zone and no $T_\text{Type}$ is available, the latter is estimated from the W2$-$W3 color (see Sect. \ref{sec:est-ttype}).
        \item If a source is placed in the upper limit zone, its $T_\text{Type}$ is not considered if available, nor calculated if not available. In this case, values calculated later in the algorithm are considered upper limits. 
    \end{itemize}
    \item The $T_\text{Type}$ of the source is used to estimate $B/T$; for sources in the upper limit zone $B/T$ is set to 1 (see Sect. \ref{sec:est-bt}).
    \item $M_\text{Bulge}$ is estimated using the previously derived $M_*$ and $B/T$ (see Sect. \ref{sec:est-stellar-bulge}).
    \item A first $M_\text{BH}$ estimate is obtained using the $M_\text{BH}-M_\text{Bulge}$ relation of \cite{Schutte2019}.
    \item The final estimate of $M_\text{BH}$ is obtained after adding $C_f$ to the first estimate. Sources with $M_\text{BH}$ in the IMBH range are removed. 
\end{itemize}
Not every source enters the algorithm. There are two main reasons for a source to be rejected during the algorithms pre-processing: (a) the source does not have a spectroscopic $z$ available, or (b) the quality of the WISE magnitudes are not considered `usable', i.e., \texttt{qph} from the AllWISE catalog is not equal to \texttt{A, B, C} or \texttt{U}, for the W1, W2 and W3 bands. 
A source can be dropped during an intermediate step of the algorithm if $M_*$ or $M_\text{BH}$ are considered outliers (described in Sects. \ref{sec:est-stellar-bulge} and \ref{sec:est-bh}).

The algorithm uses a Monte Carlo approach to estimate errors. It generates random normal samples (using arrays of size $10^4$) for the W1, W2, and W3 magnitudes and respective mean photometric errors for each source. These distributions are  propagated through the algorithm, where  errors in the scaling relations used, if available, are considered, thus delivering a final $M_\text{BH}$ (or other estimated quantity) with asymmetric error bands for each source that was not rejected. The $C_f$ is applied to the final distribution of values and not only to the nominal value. The nominal, low, and high values reported are the median and $1\sigma$ percentile values of the final distribution, respectively. Two examples of the final error distributions can be seen in Fig. \ref{fig:hist_mc}.

\section{Results}\label{sec:results}

\subsection{WISE2MBH final sample}\label{sec:output-statistics}

\begin{table*}
\centering
\caption{Statistics of the WISE2MBH final sample.}
\resizebox{\textwidth}{!}{%
\begin{tabular}{@{}lcccccccccc@{}}
    \toprule
                & \multicolumn{2}{c}{All $M_\text{BH}$} & \multicolumn{2}{c}{Galaxy} & \multicolumn{2}{c}{QSO}  & \multicolumn{2}{c}{RS} & \multicolumn{2}{c}{Unknown} \\
                & Total   & New (\%)        & Total  & New (\%)         & Total & New (\%)          & Total & New (\%)          & Total  & New (\%)             \\ \midrule
    Any         & 2010718 & 1577941 (78.48) & 586013 & 435903 (74.38)   & 15800  & 7372 (46.66)     & 6577  & 6439 (97.90)      & 1402328 & 1128227 (80.45)     \\
    Est.        & 1901419 & 1492582         & 555924 & 411201           & --     & --               & --    & --                & 1345495 & 1081381             \\
    Est. (\%)   & 94.56   & 74.23           & 94.87  & 70.17            & --     & --               & --    & --                & 95.95   & 77.11               \\
    Uplim.      & 109299  & 85359           & 30089  & 24702            & 15800  & 7372             & 6577  & 6439              & 56833   & 46846               \\
    Uplim. (\%) & 5.44    & 4.26            & 5.13   & 4.22             & 100    & 46.66            & 100   & 97.90             & 4.05    & 3.34                \\ \midrule
    Rejected    & \multicolumn{2}{c}{84869} & \multicolumn{2}{c}{--}  & \multicolumn{2}{c}{--}   & \multicolumn{2}{c}{--}   & \multicolumn{2}{c}{--}     \\ \bottomrule
\end{tabular}
}
\label{table:summary_output}
\end{table*}

\begin{figure*}
    \centering
    \includegraphics[width=\textwidth]{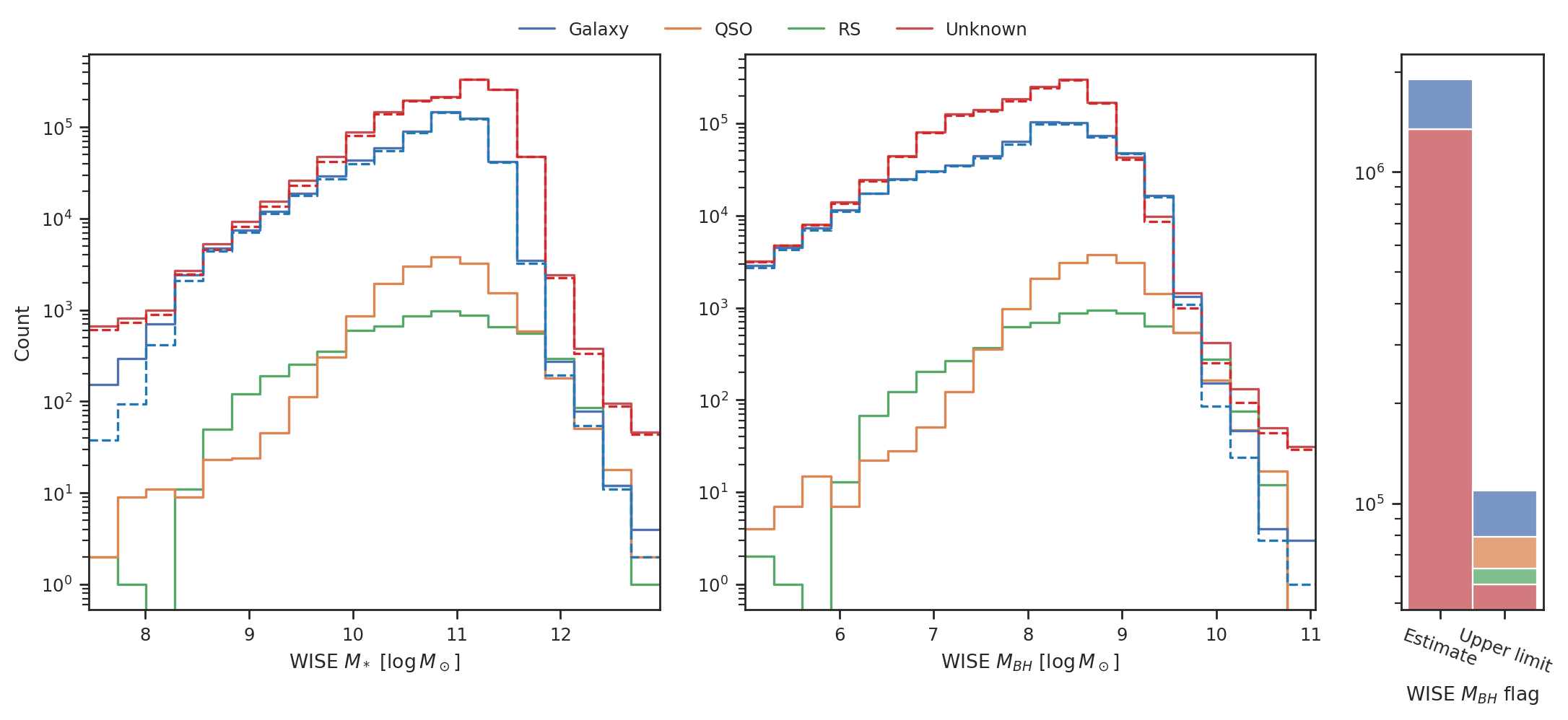}
    \caption{Distribution of $M_*$ (left panel) and $M_\text{BH}$ (middle panel) for the WISE2MBH final sample.
    For Galaxies and Unknown, two distributions are shown: the dashed line for estimates and the solid line for both estimates and upper limits.
    The distributions change slightly from one panel to the other, due to the use of $B/T$ to obtain $M_\text{Bulge}$. 
    The right panel shows a stacked histogram illustrating the number of $M_\text{BH}$ estimates and upper limits for each object type. }
    \label{fig:hist_data_output}
\end{figure*}

When used with the WISE2MBH parent sample, the algorithm generates $\sim$2 million $M_\text{BH}$ estimates and upper limits, rejecting only $3.9\%$ of the parent sample. A summary of the statistics in the final sample is shown in Table \ref{table:summary_output}. 

Percentages listed represent the percentage with respect to the total value of a given object type. New WISE2MBH estimations are those for which no previous measurement / estimate of $M_\text{BH}$ existed in the ETHER sample. Almost $80\%$ of the final sample are first-time $M_\text{BH}$ estimates or upper limits, most of which come from galaxies and unknown object-type sources that reside in the estimate zone of Fig. \ref{fig:colorcolor}.

We provide a table, in FITS format, of our WISE2MBH final sample and its derived quantities.  The table  provides the source name, RA and DEC in degrees from the AllWISE catalog, spectroscopic redshift, object type, $T_\text{Type}$ used (either from ETHER or calculated by the algorithm), plus the estimates of $M_*$, $B/T$, median plus 1$\sigma$ values of $M_\text{BH}$ as estimated by the algorithm and a quality flag for each source. An excerpt of the online table is shown in Table \ref{table:example_sources}.

The 7-digit quality flag provided in the table  stores  information useful for the selection of subsamples. The first four digits of this flag sequentially report the photometric quality of the measurements in the W1, W2, and W3 bands, as well as the extension flag of the source of AllWISE. The fifth digit serves as a binary indicator for the upper limit condition, where 0 denotes an estimate and 1 an upper limit. The sixth digit characterizes the K-correction quality; a value of 0 denotes an optimal correction, 1 is a suboptimal correction, and 2 indicates that K-correction was not applied. Lastly, the seventh digit denotes the origin of the $T_\text{Type}$ estimate. A value of 0 implies a source with a known $T_\text{Type}$, 1 indicates a $T_\text{Type}$ estimated by the algorithm, and 2 indicates that $T_\text{Type}$ was not estimated by the algorithm due to an upper limit condition. In particular, it is important to clarify that the seventh digit does not inherently rank the quality of the $T_\text{Type}$ estimate as superior or inferior, as the available $T_\text{Type}$ values can come from binary classifications, as described in \citet{Dobrycheva2013}, or consider all morphological categories, as in the case of 2MRS. This is discussed in Sect. \ref{secc:hetero}.

The highest quality sources (HQS) are classified as \texttt{AAA500} in the first six digits, while the lowest quality sources (LQS) have the fourth and fifth digits equal to \texttt{01}. Examples of HQS and LQS sources can be seen in Fig. \ref{fig:good_bad}. 
These sources illustrate the wide range of sources which the algorithm deals with, with HQS being mostly nearby galaxies large enough to have top quality in W1, W2, W3 and also being considered completely extended, while LQS are mostly QSO and compact very late-type galaxies.
While HQS is not a strict proxy of the best $M_\text{BH}$ estimates, we recommend using this sub-sample (118367 HQS, more than half of them (82812) also classified as galaxies) when a high reliability sample is required. Most QSO and RS are classified as LQS (all of their $M_\text{BH}$ are upper limits), due to the resolution limits of WISE and 2MASS, and/or AGN contamination (see Sect. \ref{sec:assum_and_lim}).

The errors in the estimates tend to be smaller in HBF sources than in LBF sources, since the extension of the sources allows for better quality in the WISE magnitudes. In Fig. \ref{fig:err_per_bf} it is possible to see that behavior for a few tens to thousands of sources at LBF.

Final distributions of $M_*$ and $M_\text{BH}$ for each object type can be seen in Fig. \ref{fig:hist_data_output}. It is clear that QSO and RS appear to have more massive $M_*$ and $M_\text{BH}$ estimates: this is due to these being upper limits (i.e. assuming $B/T=1$). Galaxies are primarily estimates: the $B/T$ ratio is used here to estimate $M_\text{Bulge}$ and $M_\text{BH}$. More than $80\%$ of sources with unknown type are estimates: their distribution is almost the same as galaxies but very different from QSO and RS. 

\begin{table*}
\centering
\caption{Excerpt of the WISE2MBH final sample.}
\label{table:example_sources}
\resizebox{\textwidth}{!}{%
\begin{tabular}{@{}llllllllllll@{}}
\toprule
Name (WISEA) & RA & DEC & $z$ & Object type & $T_\text{Type}$ & $M_*$ & $B/T$ & $_{50}M_\text{BH}$ & $_{16}M_\text{BH}$ & $_{84}M_\text{BH}$ & Quality \\
 & ($^{\circ}$) & ($^{\circ}$) & & & & ($\log M_\odot$) & & ($\log M_\odot$) & ($\log M_\odot$) & ($\log M_\odot$) & \\
 \midrule
J005254.81-260228.5 &  13.228403 & -26.041252 & 0.075047 &      Galaxy &    8.00 &  10.33 & 0.13 &    7.13 &       6.49 &       7.76 & AAA0001 \\
J085853.20+381944.2 & 134.721705 &  38.328948 & 0.089774 &     Unknown &   $-$5.0 &  10.40 & 1.00 &    8.19 &       7.58 &       8.83 & AAA0000 \\
J090734.40+502953.9 & 136.893367 &  50.498308 & 0.149000 &         QSO &  --.-- &   9.96 & 1.00 &    7.70 &       7.14 &       8.34 & AAB0122 \\
J143516.80-482147.8 & 218.820003 & -48.363285 & 0.164000 & RadioSource &  --.-- &  10.86 & 1.00 &    8.74 &       8.14 &       9.36 & AAB0122 \\
J090536.54+612254.9 & 136.402256 &  61.381936 & 0.178334 &     Unknown &    1.49 &  11.11 & 0.32 &    8.44 &       7.81 &       9.06 & AAU0001 \\
J072846.20+404954.0 & 112.192538 &  40.831674 & 0.217425 &     Unknown &    1.99 &  11.10 & 0.29 &    8.42 &       7.82 &       9.04 & AAU0001 \\
J231318.65+094133.4 & 348.327736 &   9.692628 & 0.328817 &     Unknown &    2.58 &  11.18 & 0.27 &    8.39 &       7.72 &       9.07 & AAU0001 \\
J120245.58+071133.3 & 180.689952 &   7.192605 & 0.400000 &      Galaxy &  --.-- &  10.86 & 1.00 &    8.74 &       8.06 &       9.32 & AAA0122 \\
J120939.83+344601.5 & 182.415971 &  34.767088 & 0.434100 &         QSO &  --.-- &  11.07 & 1.00 &    9.00 &       8.34 &       9.64 & AAA0122 \\
J230846.43+173546.2 & 347.193464 &  17.596178 & 0.470583 &     Unknown &    4.81 &  11.12 & 0.19 &    8.19 &       7.59 &       8.82 & AAU0001 \\
\bottomrule
\end{tabular}%
}
\flushleft\footnotesize{\textit{Note:} Some values are rounded for table presentation. The full version of this table is available online via CDS.}
\end{table*}

\subsection{Local black hole mass function}\label{sec:bhmf}

\begin{figure*}
    \centering
    \includegraphics[width=\textwidth]{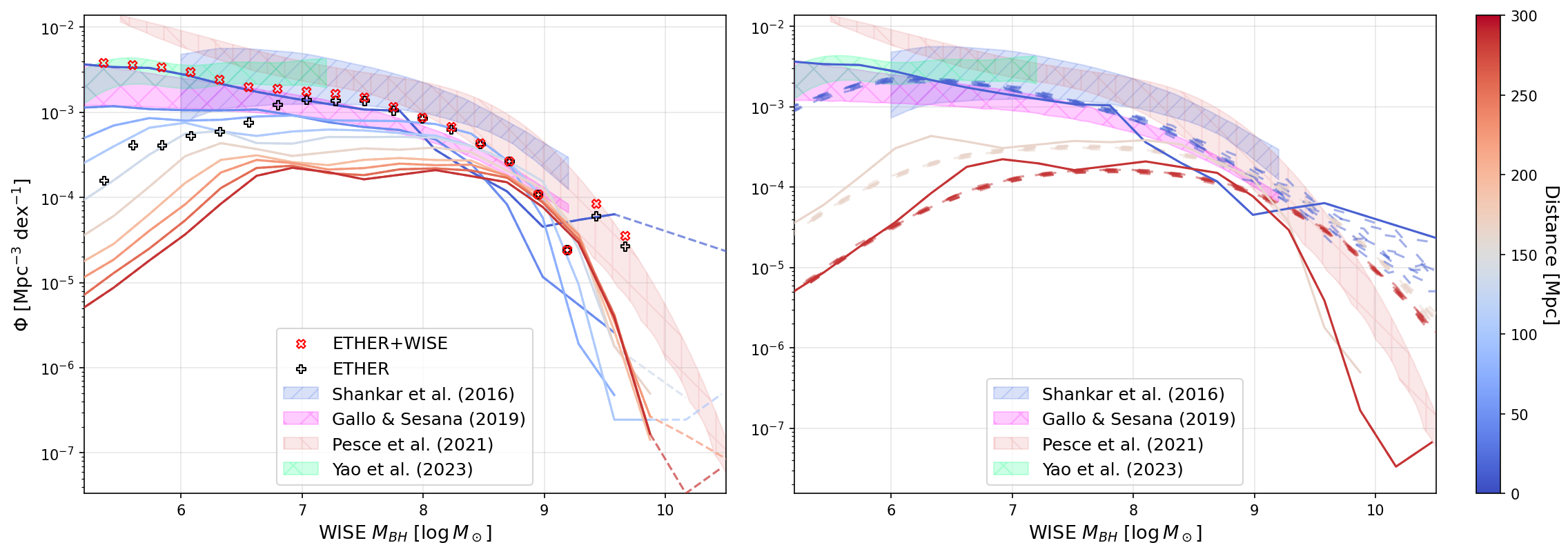}
    \caption{Left: Solid lines show the black hole mass function (BHMF) in the WISE2MBH final sample for shells of width 30 Mpc ending at distances of 30 to 300 Mpc, following the color bar at right; at high black hole masses, the BHMFs are uncertain and are shown with dashed lines. ETHER+WISE and ETHER points (see inset) represent the closest BHMF (0 $-$ 30 Mpc) present in ETHER with and without considering WISE $M_\text{BH}$ estimates. Right: Same as left, but now for three shells ending at 30, 180 and 300 Mpc, showing the BHMF using median values (solid) and ten curves showing the BHMF created using random sampling from the PDF of each $M_\text{BH}$ estimate (dashed; see text). For reference both panels show four BHMFs independently derived by \protect\citet{Shankar2016,Gallo2019,Pesce2021,Yao2023}.} 
    \label{fig:bhmf}
\end{figure*}

The local black hole mass functions (BHMFs) produced using the $M_\text{BH}$ estimated from the WISE2MBH final sample can be seen in Fig. \ref{fig:bhmf} for shells of increasing distance out to 300 Mpc. These WISE2MBH final sample BHMFs are compared to the local BHMF in \citetalias{Ramakrishnan2023}, and other  BHMFs from independent methods: \citet[compensated $M_\text{BH}$-$\sigma$]{Shankar2016}, \citet[X-ray]{Gallo2019}, \citet[$M_\text{BH}$-$M_*$]{Pesce2021} and \citet[TDE]{Yao2023}. The WISE2MBH final sample BHMF was derived using only estimates from WISE.
At $\log M_\text{BH}>10$, the WISE2MBH final sample has only a few (elliptical) galaxies, so that the derived BHMFs shows large fluctuations given the small number statistics. For this reason, the BHMFs at $\log M_\text{BH} \gtrsim10$ are shown as dashed lines.

The symbols in the figure show two cases of the BHMF for BHs at $D\leq30$ Mpc: (a) using only estimates present in ETHER before the WISE2MBH algorithm was used (black symbols), and (b) using ETHER estimates, and in the absence of these, using  estimates from WISE2MBH (red symbols). The difference between these two cases, and the WISE2MBH BHMF at $D\leq30$ Mpc (darkest blue line) is clear. At low masses ($\log M_\text{BH}\leq 7$) the ETHER-only BHMF is significantly lower than the other two. The ETHER+WISE (red symbols) and WISE2MBH BHMF agree well between each other and primarily sit on or between previously derived literature expectations. At the highest masses, the ETHER and ETHER+WISE BHMFs are above previous literature predictions: of the 17 SMBH with log $M_\text{BH}$ $\geq$ 9, the majority are M-$\sigma$ estimates with $\sigma$ from LAMOST and SDSS, so their true $M_\text{BH}$ are sensitive to both errors in $\sigma$ and in the high-mass M-$\sigma$ relationship. These BHMFs in the highest mass bins should thus be treated with caution, and will be explored in more detail in \citnp{Hernández-Yévenes et al.}. 

At the high mass end, the WISE2MBH BHMFs in all distance shells except for the closest one, drop significantly below the ETHER and ETHER+WISE BHMFs and also the literature expectations. That is, when considering median values of $M_\text{BH}$, pure WISE-based estimates do not find the few massive known black holes in the local universe. The increasing incompleteness of the WISE2MBH BHMFs with distance, especially at lower masses, is due to the redshift distribution in the sample.

Why does WISE2MBH not recover the shape of the BHMF at the massive end? We explain this as a result of not considering the intrinsic scatter of the relationship between the bulge mass and the SMBH mass. Indeed, when we consider the predicted probability density function (PDF) of each  $M_\text{BH}$ estimate from WISE2MBH (the dispersion of each PDF is dominated by the dispersion in the bulge-to-SMBH mass conversion) and randomly select a value from this PDF (instead of its median), the high-mass black hole populations are recovered and the WISE2MBH BHMFs are even higher than previous literature BHMFs (right panel of Fig. \ref{fig:bhmf}). We do the above exercise 10 times in each distance shell to create different PDF-sampled BHMFs. Even through the closest distance shell shows considerable fluctuations in its ten PDF-sampled BHMFs, all three shells consistently show BHMFs close to, but higher than, previous literature BHMFs.  

The combination of less massive $M_*$ estimates and low $B/T$ for spiral galaxies in the sample can lead to differences at the low mass end when comparing WISE2MBH BHMFs with other BHMFs in the literature.
In any case the ETHER-only BHMF is lower than literature expectations due to the construction of the sample: galaxies with an estimated black hole mass of $\log M_\text{BH}\leq5.5$ were eliminated from the sample, but note that galaxies without black hole masses remained in the sample and for these WISE2MBH could later add a mass estimate. The ETHER+WISE BHMF is in good agreement with other BHMF that obtain similar densities.

The BHMFs from \cite{Gallo2019} and \cite{Yao2023} are mostly focused in lower mass ranges. The former used the stellar mass function derived from Galaxy And Mass Assembly (GAMA) together X-ray imaging from the \textit{Chandra} Observatory, to constrain the BHMF down to low-mass regimes ($\log M_\text{BH}\geq4$). The latter was derived by studying tidal disruption events (TDE), which are less frequent in SMBH with higher masses ($\log M_\text{BH}\geq7$); they argue that their BHMF over $\log M_\text{BH}$ 5 to 7 could be considered as an upper limit. With that information, a limit was established and our BHMF is in good agreement with that.

The BHMF of \cite{Shankar2016} is the most widely used for comparisons, due to the compensated phenomenology used to derive it, as compared to the BHMF previously derived in \cite{Shankar2009}. The \cite{Shankar2016} BHMF corrects the low-mass range, now showing a clear downward trend, but it is still limited in $M_\text{BH}$ range. \cite{Pesce2021} used the \citeauthor{Shankar2009} BHMF to build an updated BHMF which extrapolates to higher mass ranges, but does not update the low-mass end BHMF. The WISE2MBH BHMFs are lower at both the high and low mass end as compared to \cite{Shankar2016} and \cite{Pesce2021}.

Given recent interest in black hole binaries (BHB) over a wide range of masses \citep[for a review, see][]{DeRosa2019}, our predicted BHMF favors recent studies. \cite{Sato2023} discussed the need for a larger population of inspiraling supermassive BHB (SMBHB) compared to current predictions to explain the stochastic gravitational wave background found by pulsar timing array (PTA) collaborations. Their prediction agrees with our closest WISE2MBH PDF-sampled BHMF; favoring the hypothesis. \cite{Izquierdo2024} predict a similar local BHMF when trying to connect LISA-detectable binaries ($\log M_\text{BH} < 7$) to SMBHB populations: our ETHER+WISE and most-local PDF-sampled BHMF presents almost perfect agreement with the predicted low-mass population in their work, while our WISE2MBH PDF-sampled BHMF is almost one dex lower at the high mass end (see their Fig. B1).

\section{Discussion}\label{sec:discussion}
\subsection{Assumptions and limitations}\label{sec:assum_and_lim}
The algorithm makes several assumptions and approximations, potentially introducing biases, particularly for certain types of galaxies. As highlighted by \citet{Jarrett2019} and \citet{Cluver2020}, despite the reduced sensitivity of the W3 band compared to W1 and W2, leading to fewer detections \citep[as noted by][]{Jarrett2013}, the color W2$-$W3 has been shown to be an effective indicator of morphology. Therefore, we choose to use this color to estimate the morphology. The galaxy-averaged W1$-$W2 color (rather than the colors of the individual components, e.g., bulge and disk) is assumed to correctly trace the averaged M/L ratio as \citetalias{Cluver2014} and \citetalias{Jarrett2023} suggest, and is effectively used for this purpose.

\begin{figure}
    \centering
    \includegraphics[width=0.48\textwidth]{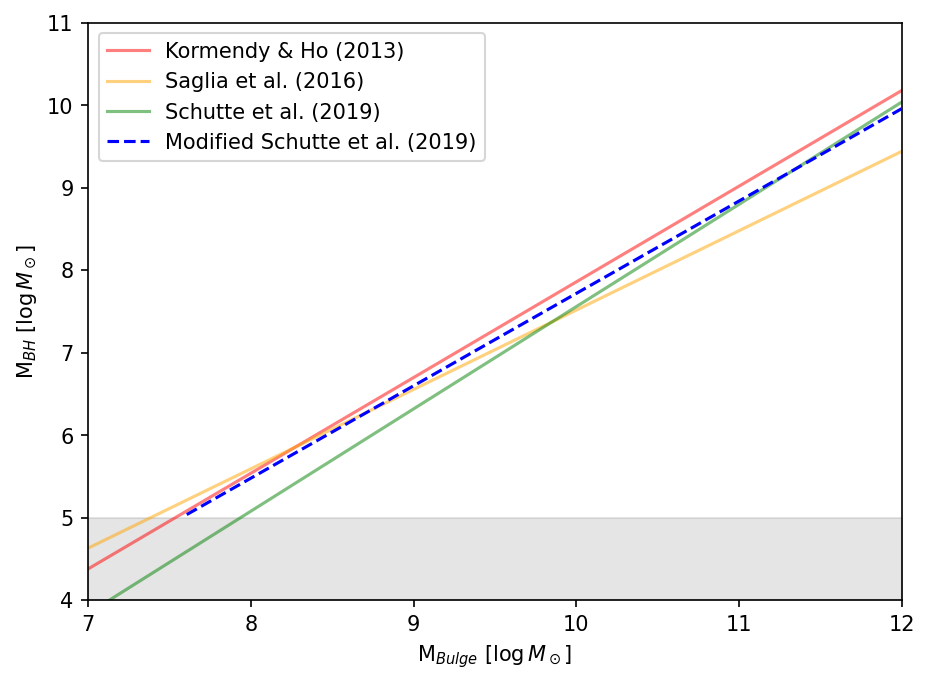}
    \caption{Comparison of different $M_\text{BH}-M_\text{Bulge}$ scaling relations from the literature, including \citet{Kormendy2013, Saglia2016, Schutte2019} and the modified scaling presented in this work. Grey area represents the limit of the WISE2MBH algorithm for $\log M_\text{BH}<5$, where it drops all estimates.}
    \label{fig:scalings}
\end{figure}

Our estimations of $M_*$ come from the method of \citetalias{Cluver2014}, which presents a different dependence of $M/L$ on W1$-$W2 compared to \citetalias{Jarrett2023}, which was the method for which the K-corrections were developed. Despite this, our most important goal with respect to the ETHER sample is not  $M_*$ estimates, but rather $M_\text{BH}$ estimates. The latter showed better agreement with measured $M_\text{BH}$ values in the literature when using the method of \citetalias{Cluver2014}. 

We initially used the empirical relationship of \cite{Schutte2019} to convert the WISE-based bulge masses to a SMBH mass, given that this is the most recent study of the relationship, and extends to relatively low black hole masses. For this conversion,  a comparison of WISE2MBH $M_\text{BH}$ to the control sample of SMBH measurements and high quality estimates shows a slight offset from equality. The best fit compensation factor ($C_f$) is positive for an initial $M_\text{BH}$ estimate with $\log M_\text{BH}<9.42$. When the first estimate of $\log M_\text{BH}$ is close to the realm of IMBH, this can lead to a large compensation factor. For example, when the first $M_\text{BH}$ estimate is $\log M_\text{BH}=4.6$, thus a $C_f=0.5$, then a final $M_\text{BH}$ estimate of $\log M_\text{BH}=5.1$ is stored. In this example, if $C_f$ was not applied, the source would have been dropped from the final sample. 

When combining the \citeauthor{Schutte2019} scaling relationship with the compensation factor (Eq. \ref{eqn:finalmbh}; see also Fig. \ref{fig:scalings}), it is clear that the final scaling relationship we derive and use lies in-between (Fig. \ref{fig:scalings}) the $M_\text{Bulge}-M_\text{BH}$ scaling relations of \cite{Kormendy2013} and \cite{Schutte2019}, being more in agreement with the former (latter) at lower (higher) masses. This shows that our final estimates effectively show a population of SMBH with masses larger than that predicted by \citeauthor{Schutte2019}. 
Given that we use only galaxies with detections in all of W1, W2, and W3, and that SMBH with final mass estimates at $\log M_\text{BH} < 5$ are deleted, the WISE2MBH algorithm is highly incomplete in estimates for dwarf galaxies, so the compensation factor and thus the final $M_\text{Bulge}-M_\text{BH}$ scaling relationship are calibrated using only more massive SMBH. 

The compensation factor obviously biases the final WISE2MBH $M_\text{BH}$ estimates to predict a population of SMBH mostly similar to the few local sources that have measurements or reliable estimates of $M_\text{BH}$. Given that we use this same $C_f$ for all redshifts in WISE2MBH, this can lead to systematic errors at higher redshifts. This underlines the need for more and better direct measurements of $M_\text{BH}$ over a range of redshifts, plus a review of systematics in previous measurements \citep[e.g.,][]{Liepold2023, Osorno2023}.

The algorithm is limited by the relatively low angular resolution of WISE ($\geq6.1\arcsec$), i.e., its limited capacity to constrain the extension or morphology of compact sources. This factor, combined with a certain degree of AGN contamination in some cases, exacerbates the classification of some targets as upper limits (rather than estimates) using the color-color criteria described in Sect. \ref{fig:colorcolor}, causing us to use unnecessarily large values of $M_\text{Bulge}$ (due to fixing $B/T=1$) and thus providing overestimated $M_\text{BH}$ values as an upperlimit. 

\protect\subsection{Using WISE2MBH at higher redshifts}

\begin{figure}
    \centering
    \includegraphics[width=0.48\textwidth]{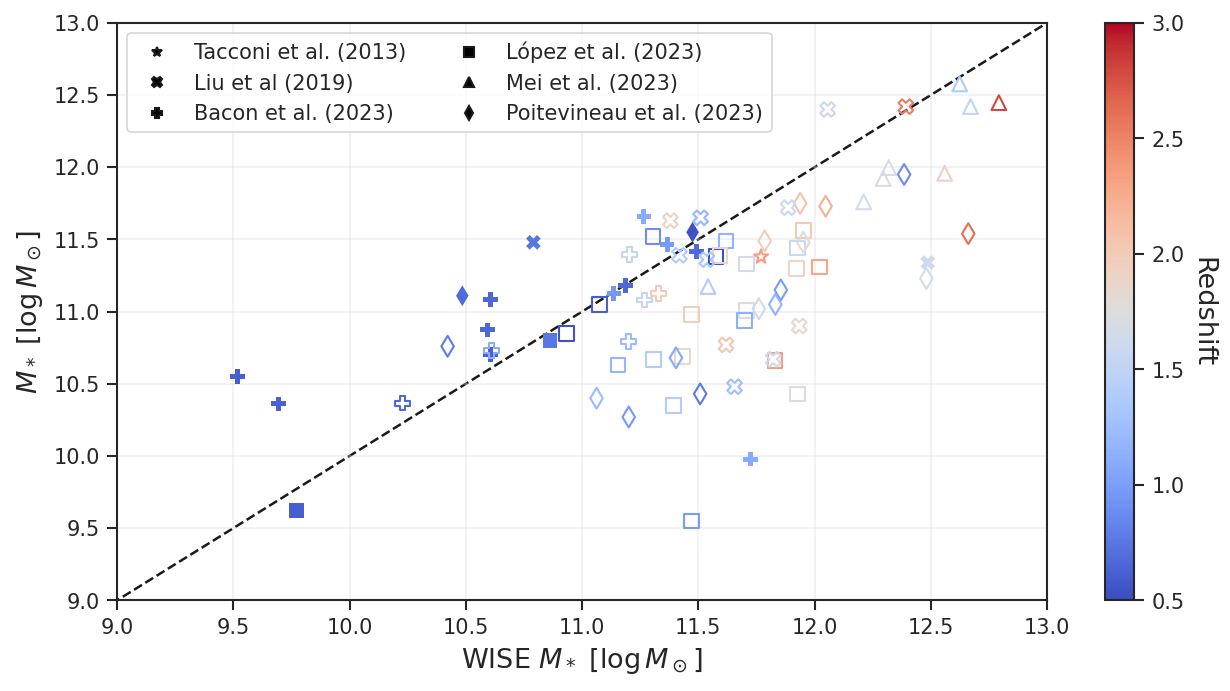}
    \caption{WISE2MBH-derived $M_*$ compared to the high redshift ($z\geq0.5$) samples of \citet{Tacconi2013,Liu2019,Bacon2023,Lopez2023,Mei2023,Poitevineau2023}, with colors specified in the inset. The black dashed line is the line of equality. Datapoints are colored by redshift following the color bar on the right (blue is $z=0.5$ and red is $z\sim3$). Filled and unfilled markers are used for estimates and upper limits  of WISE2MBH $M_*$, respectively; all literature values are estimates.}
    \label{fig:stellar_comp_hiz}
\end{figure}

In this work we have presented and discussed the use of the WISE2MBH algorithm up to redshift 0.5. This since K-corrections to the WISE photometry are relatively reliable up to this redshift, and the scaling relationships used are derived primarily in local galaxy samples. 

In principle, the WISE2MBH algorithm can be extended to the redshift range 0.5--3 by using the K-correction lookup tables from \citetalias{Jarrett2023}. However, the results over this redshift range have significantly larger uncertainties both due to the larger uncertainties in the K-corrections and due to the several other scaling  relationships used being derived and calibrated for lower redshift samples ($z \leq 0.5$). 

Specifically, the K-corrections, as presented in \citetalias{Jarrett2023}, have been made available for redshifts up to $z=3$, but have been shown to be reliable only at $z\leq0.5$. The zone selection and object criteria explained in Sect. \ref{sec:ident-sources}, have been applied systematically to every source within the WISE2MBH parent sample. Object selection and upper limit or rejection criteria are based on K-corrected fluxes and are thus more reliable at $z\leq0.5$. The $T_\text{Type}$ estimator, an essential component of the algorithm, was calibrated using the 2MRS sample, whose sources are $z\leq0.15$. Lastly, the method used to estimate $M_*$, as explained in \citetalias{Cluver2014}, has been calibrated for $z\leq0.12$. 

To test the algorithm at higher redshifts, we use WISE2MBH to derive estimates for ETHER galaxies at redshifts 0.5--3.
Figure \ref{fig:stellar_comp_hiz} compares WISE2MBH $M_*$ estimates with literature values of $M_*$ for galaxies at $z\geq0.5$ from a control sample from multiple catalogues \citep{Tacconi2013,Liu2019,Bacon2023,Lopez2023,Mei2023,Poitevineau2023}. In this figure, the filled markers show sources with WISE $M_*$ estimates and the unfilled markers show sources with WISE $M_*$ upper limits. All literature values are $M_*$ estimates.

In the context of these comparisons between estimates and upper limits against the control sample, the following observations hold:
\begin{itemize}
    \item WISE-based estimates, having undergone prior K-correction, predominantly show a slight underestimation in their  $M_*$ values.
    \item WISE-based upper limits, which did not undergo K-correction, typically show an overestimation in their WISE2MBH $M_*$ values. This tendency is consistent with the possible AGN contamination of these sources and the absence of K-correction. Since these are explicitly marked as upper limits in our algorithm, this does not lead to errors.
\end{itemize}

The WISE2MBH algorithm has been shown to be reliable in recognizing sources that are contaminated by powerful AGN emission. To err on the safe side, and given that these sources have insignificant color-color evolution across redshift \citep[e.g.,][]{Mateos2012}, we do not use K-corrections for these sources, and mark their mass estimates as upper limits.
Nevertheless, it should be noted that the algorithm has a significantly decreased accuracy in estimating $M_*$ at redshifts higher than $z=0.5$.

\subsection{Building on the algorithm and heterogeneity}\label{secc:hetero}
The WISE2MBH algorithm can be considered as an auxiliary tool for obtaining $M_\text{BH}$ estimates from a homogeneous dataset. This dataset provides consistent estimates of $M_\text{BH}$ using WISE data for most sources and a consistent set of relations, following accepted ideas and scaling relations applicable to the majority of extragalactic sources. Although the WISE2MBH final sample is heterogeneous in its composition (i.e. different extragalactic objects), users can define subsamples to recover homogeneity.

The use of a single, well-calibrated dataset from observations with the same instrument and method can ensure consistency and comparability of results, but  may limit their generalizability and bias the final conclusions. The process described in this work does not necessarily rely solely on WISE data. WISE is used to obtain physical properties of extragalactic sources in the ETHER sample, but external data, such as source classification and morphological types, are also employed in some cases.

Independently derived physical properties, such as $M_*$ or $M_\text{Bulge}$ can be input to an intermediate stage of WISE2MBH in order to obtain $M_\text{BH}$ estimates. This approach injects heterogeneity into the algorithm and its results.

Despite the use of a homogeneous parent sample, both $M_\text{BH}$ (low-z) and  $M_*$ (low- and high-z) were compared with heterogeneous control samples, the former being more important to discuss. The $M_\text{BH}$ control sample is described in Table \ref{table:control_sample}. The primary methods used to measure $M_\text{BH}$ in the control sample are stellar dynamics, gas dynamics, and reverberation mapping (RM); all known for their reliability and precision in measuring $M_\text{BH}$. Poor quality $\sigma$, single-epoch RM, and other methods used to obtain $M_\text{BH}$ in \citetalias{Ramakrishnan2023} were excluded. Despite the heterogeneity of the control sample, the ratio between the literature measured or estimated $M_\text{BH}$ and the WISE2MBH $M_\text{BH}$ estimate was calculated in every case, and no prominent differences or scatter was found. Furthermore, the WISE-based bulge luminosity to black hole mass  scaling closely follows the relationship  of \cite{Kormendy2013}, which is widely used in the literature. 

Authors who wish to work with a completely homogeneous subsample of the WISE2MBH parent sample which is based only on WISE data and spectroscopic redshift, could define the subsample as follows: (a) only sources with object type galaxy or unknown, and (b) only sources with a quality flag ending in 2. These constraints ensure that authors work with estimates that only used WISE data for the classification of upper limits and omit estimates that made use of previously known $T_\text{Type}$.

\subsection{Relevance for the EHT and ngEHT}\label{sec:relevance-ngeht}
The EHT (and future ngEHT) is the best facility for the imaging of the innermost environments of black holes in terms of sensitivity and resolution \citep[$\sim$10 mJy and $\sim$15 $\mu$as,][]{Doeleman2019, Pesce2022, Johnson2023, Doeleman2023} for the next decade, opening the possibility of imaging (and making movies of) tens to hundreds of SMBH in the nearby Universe. Relevant science goals include testing general relativity, the role of magnetic fields in black hole accretion and jet formation. Recently, \cite{Pesce2021} have demonstrated that with current EHT facilities (at 230GHz), we can expect to resolve $\sim$5 new SMBH shadows, while with ngEHT observing at 345GHz, this number can be increased by factor $\sim$3. The challenge is to identify these very rare sources. 

In the context of scientific exploitation of the EHT, \citetalias{Ramakrishnan2023} have developed the ETHER sample and database. Combining $\gtrsim$3M sources in a parent sample of galaxies and AGN, comprehensive radio to X-ray observed spectral energy densities (SEDs), and jet and accretion flow SED modelling to predict the expected EHT flux, ETHER can provide target samples for the EHT for any given science goal. WISE2MBH was originally developed to fill large gaps in the ETHER sample: delivering both black hole mass estimates and upper limits, and galaxy morphologies. Its accuracy and ease of use, combined with its relevance to galaxy evolution studies, especially at high redshift, motivate its publication as a separate entity from ETHER.

This algorithm does not necessarily intend to replace previous estimates of $M_\text{BH}$ in the literature (except for values based on poor quality $\sigma$ or relatively unreliable 'fundamental planes'), but rather to increase the completeness of the ETHER sample.  As more precise estimations or measurements become available \citep[e.g., from SDSS Black Hole Mapper,][Sect. 2.2]{Kollmeier2017}, WISE $M_\text{BH}$ estimates can be replaced in ETHER.

In \citetalias{Ramakrishnan2023}, the authors detail a methodology that requires $M_\text{BH}$ estimates (or upper limits) and X-ray flux data, primarily from the \textit{Chandra} Source Catalog (CSC) to predict the EHT-observable flux of an SMBH. With the release of the eROSITA all-sky survey \citep[eRASS,][]{Merloni2024} DR1 an estimated $\sim1$ million new hard X-ray flux measurements have become available. Integrating eROSITA, WISE2MBH and ETHER, will allow SED fitting, thus radio flux estimates, for almost all sources in eROSITA. This is a critical step for target selection for the ngEHT.

\section{Summary}\label{sec:conclusion}
This work presents a simple and new algorithm to obtain stellar masses ($M_*$), morphological types ($T_\text{Type}$), bulge-to-total ratios ($B/T$), and black hole masses ($M_\text{BH}$) estimates of a galaxy. The WISE2MBH algorithm is publicly available at GitHub\footnote{A general use version of the algorithm is available at the following GitHub repository: \href{https://github.com/joacoh/wise2mbh}{https://github.com/joacoh/wise2mbh}}. This algorithm, which only requires WISE catalog data, classifies sources as galaxies or QSOs, and estimates multiple physical quantities such as $M_*$ and $M_\text{BH}$. The algorithm uses previously derived scaling relations and our own derived relations to obtain a final $M_\text{BH}$ estimate or upper limit. Using a parent sample of $\sim$2.1 million sources from the ETHER sample post-cross-match to the AllWISE and  WISE extended source (WXSC) catalogues, a final sample of $\sim$1.9 million $M_\text{BH}$ estimates and $\sim$100 thousand upper limits were calculated. Among the estimates (i.e. not considering upper limits), $\sim$$78.5\%$ are first-time estimates of known galaxies or unclassified sources. QSOs and radio sources (RS), as classified by NED, are also part of the sample, but due to the nature of their emission and the quality or extension of these sources in WISE, all of their final values of $M_\text{BH}$ are marked as upper limits. The final sample table is available online via CDS.

The detailed manual morphological classifications ($T_\text{Type}$) of galaxies in the 2MASS redshift survey (2MRS) were used to derive a relation between $T_\text{Type}$ and  WISE W2$-$W3 color, with the objective of estimating $T_\text{Type}$ for sources that do not have one previously assigned in the literature. All available and estimated $T_\text{Type}$ are used to obtain $B/T$ using an exponential relation described in Sect. \ref{sec:est-bt} that is consistent with previous studies. The obtained $B/T$ are used to calculate $M_\text{Bulge}$ from a WISE derived $M_*$. 
Finally, we use our $M_\text{BH}-M_\text{Bulge}$ scaling relation - which lies between accepted literature scaling relationships - to estimate
$M_\text{BH}$.
All uncertainties are propagated through the algorithm using a Monte Carlo approach, delivering the $1\sigma$ (upper and lower) errors of the final distributions as low and high values, respectively.

The $M_\text{BH}$ estimates were compared to a control sample of $M_\text{BH}$ measurements and reliable estimates, showing a significant difference in the linear regression analysis with respect to the equality line, i.e., an offset that causes some values to be overestimated and others underestimated. To compensate for this offset, we implement a compensation factor ($C_f$) derived with the use of the control sample. After compensation, the comparison achieves a Spearman score of $\sim0.78$ and a RMSE of $\sim0.63$. The final scaling relationship used (eqn. \ref{eqn:finalmbh}) lies between the relationships of  \cite{Kormendy2013} and \cite{Schutte2019}, being more consistent with the former at low SMBH masses and the latter at high SMBH masses. 
The mean uncertainty was calculated for the $M_\text{BH}$ estimates, considering a simple mean between low and high errors and then taking the mean value of the distribution of means, obtaining a value of $\sim0.5$ dex, showing more scatter in low bulge fraction (LBF) sources, compared to high bulge fraction (HBF) sources (see Fig. \ref{fig:err_per_bf}). 

The black hole mass function (BHMF) of the WISE2MBH final sample is in good agreement with other previously and independently derived BHMFs. The ETHER-only sample has few low mass estimates ($\log M_\text{BH}\leq6$), while the WISE2MBH final sample can provide this population of sources, and the overall combination of both samples generates the most complete BHMF. We find evidence that high mass SMBH are more common than predicted by literature BHMFs.  

When using the WISE2MBH algorithm or the final sample described in this work, it is important to take into account the assumptions and associated limitations. The algorithm provided on GitHub can be easily modified to change the scaling relations used or incorporate new ones, tailored to the user's requirements.  

Regarding the final sample, we recommend not considering all $M_*$ or $M_\text{BH}$ estimates if the main goal is to study only a few sources and/or restrict the sample to only high quality sources (HQS). In case of population studies, almost the complete final sample can be used, depending on the redshift limit, quality flag, and the requirements of the user.

The final sample was generated in a homogeneous manner, i.e. all estimates come from relations that make use of WISE cataloged data to derive physical quantities, except for the use of $T_\text{Type}$ from the literature in some cases. This gives confidence that the derived values are consistent from one to another and no externally derived physical parameters were used to obtain the final $M_\text{BH}$.

The WISE2MBH final sample is already incorporated into the ETHER sample, providing almost 3 million new $M_\text{BH}$ estimates and upper limits that, and it will be used iteratively to provide up-to-date values in case new sources are ingested into the ETHER sample. These estimates are crucial for the selection of samples of interest for the Event Horizon Telescope (EHT) and the next-generation EHT (ngEHT), and are used on each update of the sample. Its high percentage of success in estimating a new $M_\text{BH}$, combined with spectral fitting of accretion and jet models to hard X-ray data from Chandra, XMM, and eROSITA, allows one to predict radio fluxes from the accretion inflow and jets, and thus obtain a first selection of sources detectable with the EHT or ngEHT. 

\section*{Acknowledgements}
We thank Yuri Kovalev, Angelo Ricarte, Dominic Pesce, and Priyamvada Natarajan for useful discussions and feedback, and Yuhan Yao for providing black hole mass functions for comparison. We acknowledge funding from ANID Chile via Nucleo Milenio TITANs (Project NCN2023{\_}002), Fondecyt Regular (Project 1221421) and Basal (Project FB210003). T.H.J. acknowledges support from the National Research Foundation (South Africa).
This research has used the VizieR catalog access tool, CDS, Strasbourg, France. 
We acknowledge the usage of the HyperLeda database (http://leda.univ-lyon1.fr). 
This research made use of the following software: Pandas \citep{pandas}, Astropy \citep{astropy}, Numpy \citep{numpy}, Scipy \citep{scipy}, Matplotlib \citep{matplotlib}, Seaborn \citep{seaborn}, StatsModels \citep{statsmodels} and Topcat \citep{topcat}.

\section*{Data Availability}

The WISE2MBH algorithm and the K-correction tables from \citetalias{Jarrett2023} for eliptical, lenticular and spiral morphologies, can be found in the GitHub repository. The WISE2MBH final sample, which includes all the data described in Table \ref{table:summary_output}, is distributed through CDS.




\bibliographystyle{mnras}
\bibliography{main.bib} 

\appendix
\section{Diversity of objects in the WISE2MBH final sample}
\begin{figure}
    \centering
    \includegraphics[width=0.48\textwidth]{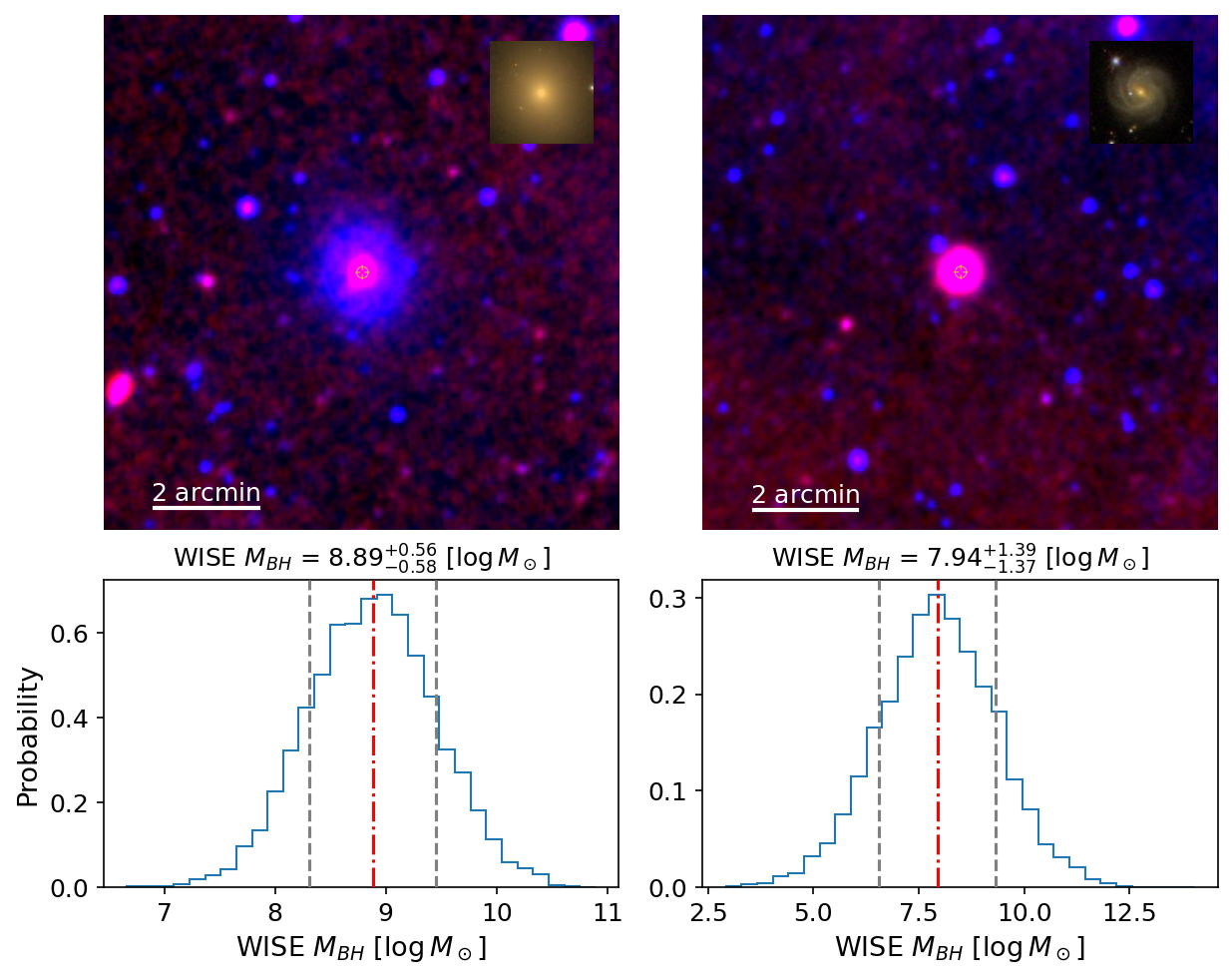}
    \caption{Top: WISE two-color images with FOV of 10' of NGC 7626 (left; a HBF galaxy) and NGC 7773 (right; an LBF galaxy): blue represents the W2 band and red the W3 band. The W2$-$W3 color clearly distinguishes between the HBF and LBF galaxies. In each panel, the SDSS DR16 image of the galaxy is shown as an insert in the upper right corner for reference. Bottom: The corresponding WISE2MBH $M_\text{BH}$ probability density function (PDF) provided by our algorithm for each source in the top row. The red vertical line denotes our final (median) $M_\text{BH}$ value and the dashed vertical lines represent the $1\sigma$ of the distribution; the reported values for the lower and upper values of $M_\text{BH}$. The HBF galaxy has smaller $M_\text{BH}$ uncertainties as compared to the LBF galaxy; a trend seen in general for LBF galaxies, e.g., Fig. \ref{fig:err_per_bf}.}
    \label{fig:hist_mc}
\end{figure}

\begin{figure}
\centering
\includegraphics[width = .48\textwidth]{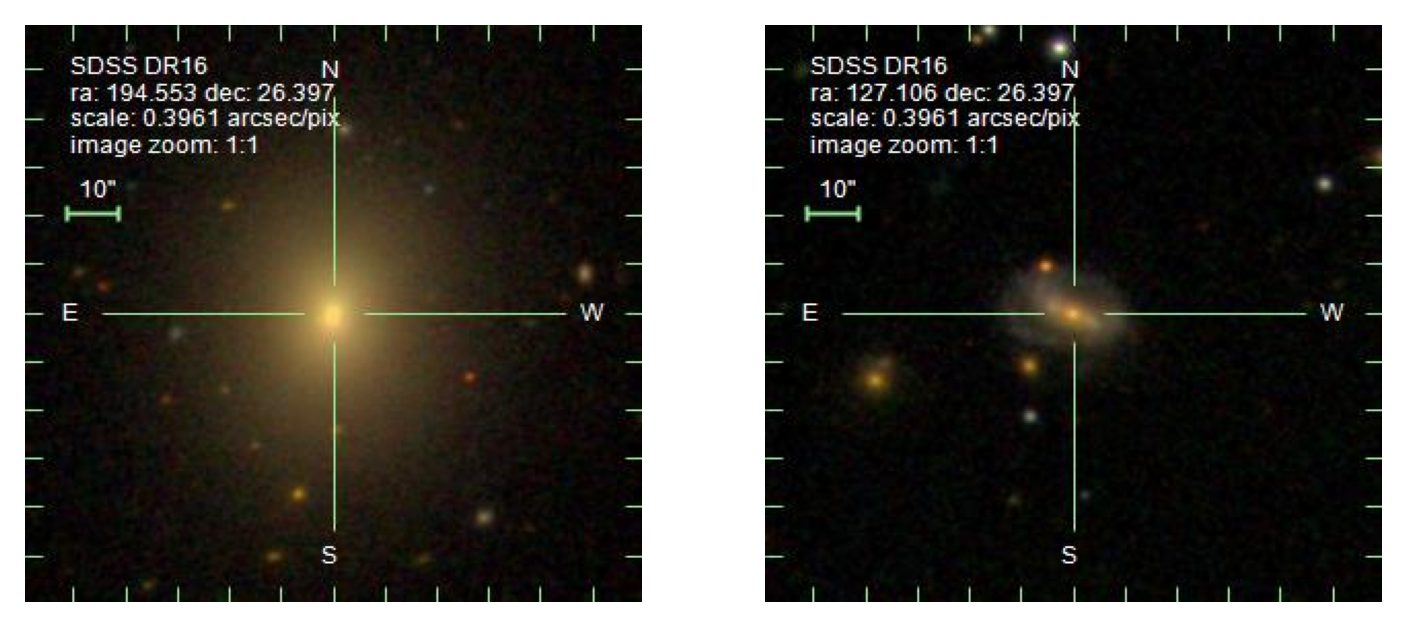}
\includegraphics[width = .48\textwidth]{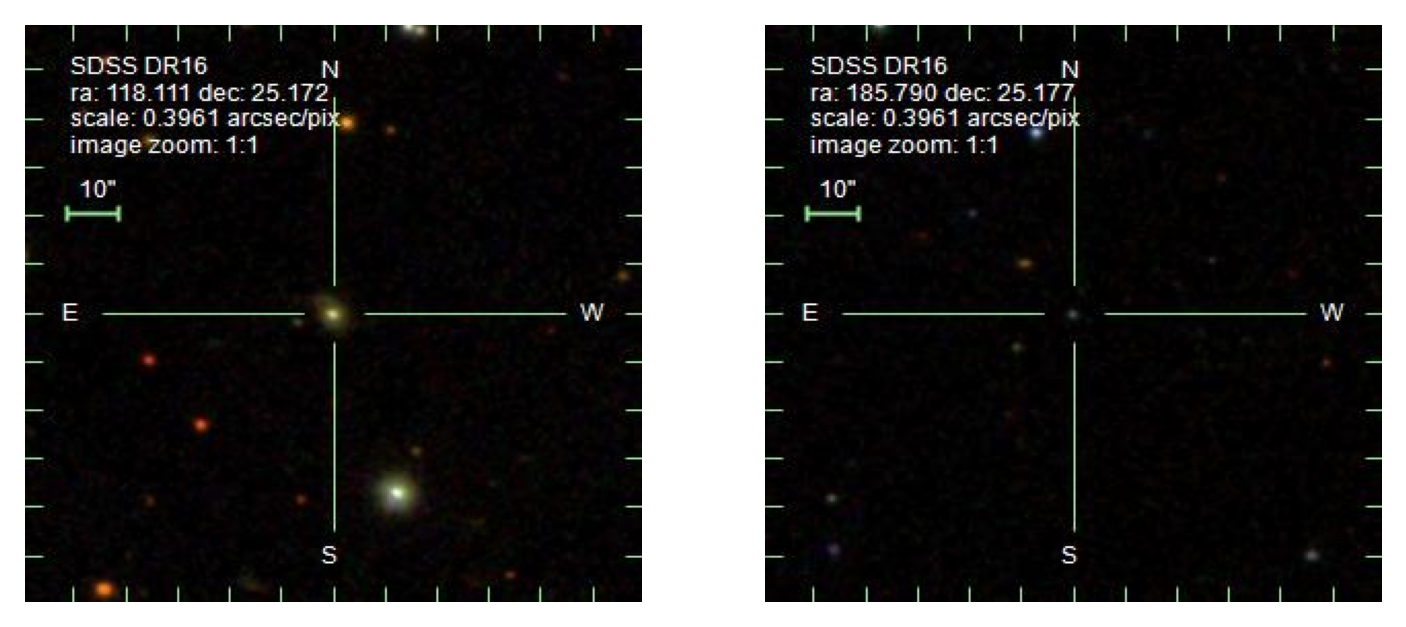}
\caption{SDSS DR16 images of HQS (top) and LQS (bottom) examples. From left to right, at the top are NGC 4849 (HBF) and SDSS J082825.42+262350.4 (LBF). At bottom are SDSS J075226.52+251020.1 and SDSS J122309.61+251036.7, both are have final $M_\text{BH}$ upper limits.}
\label{fig:good_bad}
\end{figure}

As stated in Secs. \ref{sec:ident-sources}, \ref{sec:est-ttype} and \ref{sec:output-statistics}, the object type is an optional input to the algorithm. 
The algorithm estimates multiple parameters or classifications for the objects, e.g.,  high and low bulge fractions (HBF and LBF) and high and low quality sources (HQS and LQS). An example 
comparing an LBF and a HBF object is shown in Fig. \ref{fig:hist_mc}. Figure \ref{fig:good_bad} contrasts two examples of HQS sources (top row) with two examples of LQS sources (bottom row).

\section{Comparison for various samples in W2-W3 color to T-type conversion}\label{sec:appen1}

Our conversion of W2$-$W3 color to $T_\text{Type}$ was calibrated with galaxies in the 2MRS sample (Sect. \ref{sec:est-ttype}). To explore the reliability and errors of this conversion, we used the same method of Sect. \ref{sec:est-ttype}, but for other samples present in Hyperleda \citep{Makarov2014}.

\begin{figure}
    \centering
    \includegraphics[width=0.48\textwidth]{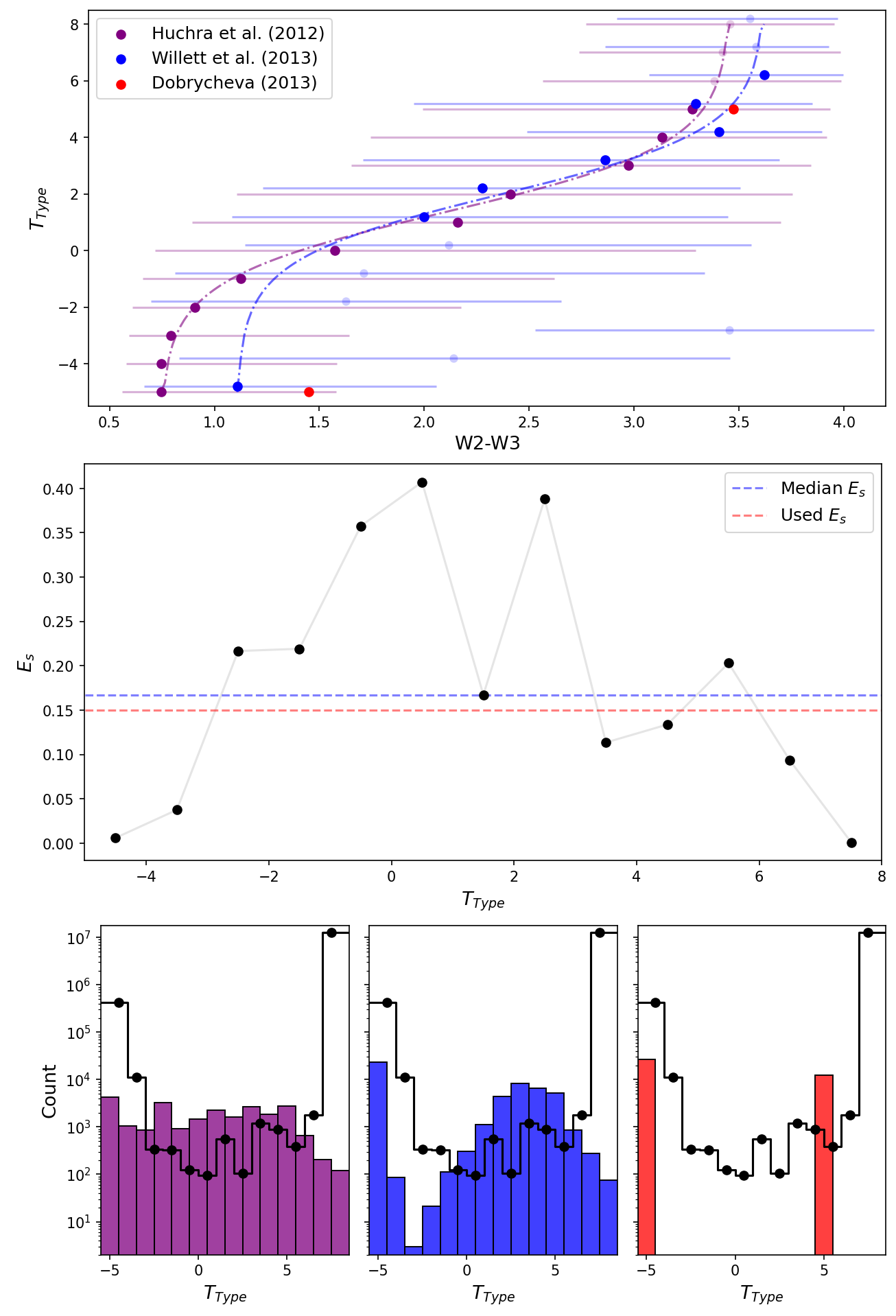}
    \caption{Top: As in Fig. \ref{fig:logit} but for all samples described in Appendix \ref{sec:appen1}. Blue dots are slightly shifted in the Y-axis to distinguish between error bars. Colors follow the legend in the top left of the panel. Middle: Effect size ($E_s$) for every consecutive distribution of W2$-$W3 color, following the order of $T_\text{Type}$. The position on the X-axis is the middle value between consecutive $T_\text{Type}$. Bottom: $T_\text{Type}$ distributions of the three samples shown in the top panel are shown in colored histograms of the corresponding color. 
    Black connected dots denote the sample sizes ($N$) required to establish a distinction between consecutive $T_\text{Type}$ bins, as derived from the medians of the 2MRS W2$-$W3 distributions in each $T_\text{Type}$ bin.}
    \label{fig:appen_figure1}
\end{figure}

Hyperleda includes the \citet[2MRS]{Huchra2012}, \citet[GZ2]{Willett2013}, and \cite{Dobrycheva2013} samples, which collectively represent more than $70\%$ of the available values for $T_\text{Type}$ in Hyperleda (25k, 310k, and 350k values, respectively). Each sample was tested following the same approach described in Sect. \ref{sec:est-ttype} for 2MRS, with results shown in Fig. \ref{fig:appen_figure1}.

The GZ2 sample exhibits a clear bimodality in T (bottom row of Fig. \ref{fig:appen_figure1}). The authors provide detailed classifications for spiral galaxies, allowing clear differentiation from Sa to almost Sd classifications (1 to 6 in $T_\text{Type}$). However, in the morphological range from lenticular to elliptical, the level of detail is completely lost: all except 530 (i.e. 98\%) of these sources are classified as $T_\text{Type}=-5$. 

\citeauthor{Dobrycheva2013} classified their sample galaxies into two bins: $T_\text{Type}$ equal $-$5 and 5  (i.e. ellipticals and spirals; bottom row of Fig. \ref{fig:appen_figure1}). 
Although this binary classification has demonstrated efficacy for machine learning training and subsequent classification of different samples \citep[e.g.,][]{Vavilova2021, Vavilova2022}, it does not provide the level of detail required for our W2$-$W3 color to $T_\text{Type}$ calibration. Nevertheless, we use this sample here only for comparison purposes. 

In contrast, the 2MRS sample shows a relatively smooth and well populated distribution of $T_\text{Type}$ values (bottom row of Fig. \ref{fig:appen_figure1}).

The distributions (median values and $1\sigma$ intervals) of W2$-$W3 for each $T_\text{Type}$ in 2MRS and GZ2 are presented in the top panel of Fig. \ref{fig:appen_figure1}. For \citeauthor{Dobrycheva2013}, only the medians are shown for comparison. 
The GZ2 fit shows a similar trend to 2MRS in the range of $T_\text{Type}$ 0 to 5, covering the same range but with smaller error bars. At the upper end, the fits differ by approximately 0.2 in W2$-$W3, which has a negligible impact on the final $M_\text{Bulge}$ estimates when comparing the GZ2 fit to 2MRS fit. At the lower end, there is a difference of almost 0.4 in W2$-$W3, leading to significant variations in the estimated $B/T$. For a source with a W2$-$W3 color of 1, the 2MRS fit gives a $B/T\sim0.5$ estimate, while the GZ2 fit results in a $B/T\sim1$, corresponding to a difference of 0.3 dex in the $M_\text{Bulge}$ estimates, which are the first estimates affected by the $B/T$ value. These systematic changes in fit have significant implications for the final estimates and the overall statistics of the sample. 
Due to the lack of detail and biased representation, the GZ2 sample and its fit were discarded from our analysis.
For the \citeauthor{Dobrycheva2013} sample, elliptical galaxies exhibit distinct shifts compared to both 2MRS and GZ2, primarily due to the binarity of the classification used.

Therefore, it is crucial to use a sample that is large enough to accurately discriminate between $T_\text{Type}$ and then to rely on the trend shown by that specific sample. 
We defined this `large enough' sample size to be $N$, and determined it to be approximately $N\sim700$ per bin by statistical power analysis, which is a widely used statistical tool for sample size determination in meta-analyses \citep[e.g.,][]{Borenstein2011,Grundler2022}. 
In statistical power analysis, the three parameters to be set are statistical power, significance threshold, and effect size. The statistical power ($P$) is often defined as the probability that a study can correctly detect a real effect (i.e., probability of avoiding a Type II error). The significance threshold ($\alpha$) is defined as the highest level of acceptable risk in terms of incorrectly rejecting a null hypothesis that is actually true (i.e., Type I error probability). Effect size ($E_s$) is a standardized technique available to measure the expected strength of the results in a study, regardless of the sample size. This $E_s$ can only be calculated for two samples, so in multisample scenarios this has to be calculated for each pair of samples in a predefined order, e.g., if the samples represent an evolution from 0 to 10 in a quantity, the pairs to calculate $E_s$ must be 0$-$1, 1$-$2, and so on.

For this analysis, we used the median values of the W2$-$W3 distribution for each $T_\text{Type}$ value in 2MRS, with parameters; $P=0.8$, $\alpha=0.05$ and $E_s=0.15$. However, the value of $E_s$ varies for each pair or consecutive distribution of W2$-$W3 following the order of $T_\text{Type}$, as shown in the middle plot of Fig. \ref{fig:appen_figure1}. The median value of $E_s$ was approximately 0.17, which corresponds to $N\sim550$. To ensure stricter statistical power, we reduced $E_s$ to 0.15. For the lower and upper ends of $T_\text{Type}$, $E_s$ can go as low as 0.001, requiring a sample size of $N\geq10^7$ to confidently claim that the consecutive distributions are, in fact, two distinct populations and not two samples from the same parent sample. However, the chosen value of $E_s$ is sufficiently strict to clearly discriminate between each distribution of W2$-$W3 for $T_\text{Type}$ ranging from -3 to 6.

In the bottom plots of Fig. \ref{fig:appen_figure1} one can see that the required $N$ (black dots)  is surpassed in most cases by the three samples (colored histograms), but notoriously larger samples are needed for the most extreme values of $T_\text{Type}$. It is clear that the results obtained from the 2MRS statistical power analysis are not directly applicable to limit the use of other samples, but similar median values are expected for the whole population of galaxies and the different $T_\text{Type}$ between samples, thus needing similar $N$, independent of the sample used.

We decided to use 2MRS over the other samples tested due to the completeness in the lenticular-spiral regime and since both 2MRS and GZ2 samples showed similar behavior in the late-type regime, even though 2MRS is almost 10 times smaller in overall sample size. 

\section{Correlations and scatter between WISE2MBH and SDSS estimates of stellar mass}\label{sec:appen3}

\begin{figure*}
    \centering
    \includegraphics[width=\textwidth]{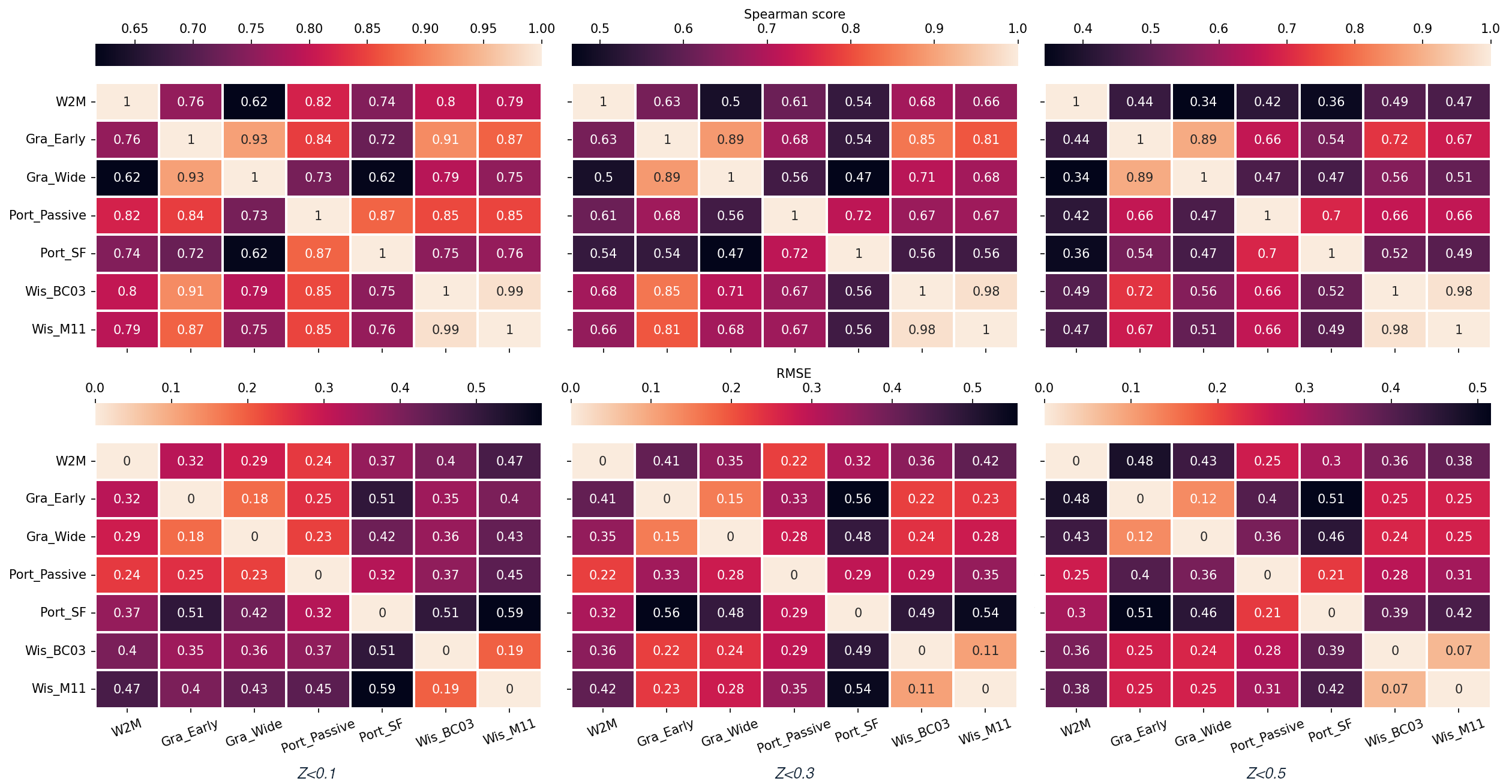}
    \caption{Correlation (top) and scatter (bottom) matrices when comparing WISE2MBH $M_*$ estimates with those from \citet[\texttt{Gra\_Early} and \texttt{Gra\_Wide}]{Montero2016}, \citet[\texttt{Port\_Passive} and \texttt{Port\_SF}]{Maraston2011} and \citet[\texttt{Wis\_BC03} and \texttt{Wis\_M11}]{Chen2012} for three redshift ranges of $z<0.1$ (left), $z<0.3$ (middle) and $z<0.5$ (right). Boxes are colored following the colorbar on top of each panel; for the scatter matrices (bottom row) the colorbar is inverted for easier understanding.}
    \label{fig:corr_scatter}
\end{figure*}

Since deriving $M_*$ is the first step towards estimating $M_\text{BH}$, the WISE-estimated $M_*$ must be accurate as compared to estimates obtained with different homogeneous methods applied to very large samples. We found that using only the \citetalias{Cluver2014} method, or only the \citetalias{Jarrett2023} method, did not lead to optimal results when comparing to literature values. Instead combining the $M_*$ estimation from \citetalias{Cluver2014} with prior K-corrections derived by \citetalias{Jarrett2023} gave the most accurate results. 

To test for bias and consistency, we compare the WISE2MBH derived $M_*$ estimates to those of three different value added catalogues (VACs) of $M_*$ available via SDSS: Portsmouth group \citep{Maraston2013} who used SED fitting with passive and star-forming templates with the Kroupa initial mass function \citep[IMF,][]{Kroupa2011}, Wisconsin group \citep{Chen2012} who used the principal component analysis (PCA) method in the optical rest-frame spectral region (3700$-$5500 \AA) with two different single stellar population models from \citet[BC03]{Bruzual2003} and \citet[M11]{Maraston2011}, and finally the Granada group \citep{Montero2016} who used the flexible stellar population synthesis (FSPS) method for early and wide formation times.

The correlation and scatter matrices considering all catalogues mentioned above and WISE2MBH for three different redshift bins are presented in Fig. \ref{fig:corr_scatter}.

For correlations (upper row of Fig. \ref{fig:corr_scatter}), Spearman scores range from $\sim0.6-0.9$ for almost every method considered in the lowest redshift bin ($z<0.1$). When considering a larger redshift bin ($z<0.3$), all correlation scores decreased, particularly for methods that take into account completely different methodologies, such as \texttt{Port\_SF} and \texttt{Gra\_Early}. When considering the entire redshift range ($z<0.5$), the scores decreased once again, now down to $0.34$ for the case of WISE2MBH and \texttt{Gra\_Wide}. 

The worst scores are found between \texttt{Gra\_Wide} and \texttt{Port\_SF}
in the $z<0.1$ (score = $0.62$) and $z<0.3$ (score = $0.47$) bins. \texttt{Port\_SF} performs the worst in intercomparisons in the $z<0.3$ bin. WISE2MBH performs well compared to SDSS VACs except in the highest ($z<0.5$) redshift bin.

When considering the scatter between multiple methods (lower row of Fig. \ref{fig:corr_scatter}) WISE2MBH shows the lowest RMSE across all the methods and redshift bins, i.e., it is the method which overall most closely agrees with all VACs on average. For WISE2MBH and \texttt{Port\_Passive}, the RMSE is $0.24$ for $z<0.1$ and even lower ($0.22$) when considering the entire redshift range. However, for WISE2MBH and \texttt{Gra\_Early}, the RMSE increases with redshift.

For the specific case of WISE2MBH and \texttt{Gra\_Early}, both correlation and scatter got worse with increasing redshift. The correlation starts reasonably well  (score = $0.76$) in the lowest redshift bin, then becomes one of the worst scores in the entire $z<0.5$ bin (score = $0.44$). The same pattern is seen in the scatter, which increases from RMSE of $0.32$ to $0.48$, the latter being one of the highest RMSE in the entire matrix. In Fig. \ref{fig:stellar_comp} it is clear that there is a systematic shift with redshift that causes the low correlation score. It is the only pair that shows this strong tendency to get worse in both metrics with increasing redshift.








\bsp	
\label{lastpage}
\end{document}